\documentclass[aps,prd,superscriptaddress,11pt,showpacs,notitlepage]{revtex4}
	\usepackage{graphicx}
	\pdfoutput=1
	\usepackage{amsmath,amssymb,mathrsfs}
	\usepackage[colorlinks=true, a4paper=true, pdfstartview=FitV,
linkcolor=blue, citecolor=blue, urlcolor=blue]{hyperref}

\usepackage{epsf}
\usepackage{epsfig}
\usepackage[usenames,dvipsnames]{xcolor}

\newcommand{\be}{\begin{equation}}
\newcommand{\ee}{\end{equation}}
\newcommand{\bea}{\begin{eqnarray}}
\newcommand{\eea}{\end{eqnarray}}

\newcommand{\bb}{\bibitem}
 
\newcommand{\eqn}{\begin{eqnarray}}
\newcommand{\eqnx}{\end{eqnarray}}

\numberwithin{equation}{section}

\begin{document}

\title{The superpotential method in cosmological inflation}

\author{C. Adam}
\affiliation{Departamento de F\'isica de Part\'iculas, Universidad de Santiago de Compostela and Instituto Galego de F\'isica de Altas Enerxias (IGFAE) E-15782 Santiago de Compostela, Spain}
\email{adam@fpaxp1.usc.es}
\author{D. Varela}
\affiliation{Departamento de F\'isica de Part\'iculas, Universidad de Santiago de Compostela and Instituto Galego de F\'isica de Altas Enerxias (IGFAE) E-15782 Santiago de Compostela, Spain}

\begin{abstract}
Scalar field cosmological inflation has a first integral relating the Hubble function and the lagrangian of the scalar field(s), which is known under the names of "Hamilton-Jacobi approach" or "superpotential equation". Here we exploit the simplicity of this superpotential equation and use it as an alternative but equivalent cosmological evolution equation during inflation, replacing the Friedman-Robertson-Walker (FRW) equations. It turns out that all inflationary observables can be calculated directly from its solution (the superpotential). Further, the superpotential equation allows for a simple and direct calculation of the slow-roll expansion to arbitrary order and, in many cases, for an exact determination of the slow-roll attractor. It also allows for a power series expansion in the inflaton field which permits to estimate the radius of convergence of the slow-roll expansion. We consider several examples of single-field inflationary models to demonstrate the simplicity and usefulness of the method.
\end{abstract}

\maketitle

\section{Introduction}
Cosmological inflation  consists in the assumption that the very early universe underwent a phase of very fast (in many scenarios, approximately exponential) expansion \cite{starob1}, \cite{guth1}, \cite{linde1} (for detailed introductions into inflation see, e.g., \cite{Dod} - \cite{Baumann}). Inflation was originally conceived to explain the absence of magnetic monopoles (predicted by many Grand Unified Theories) in the observable universe and its almost perfect (spatial) flatness. Today, there are two particularly convincing arguments in favor of some version of inflation. On the one hand, inflation provides a natural and plausible explanation for the spatial homogeneity of the visible universe, by bringing different regions of the universe in causal contact at sufficiently early times. On the other hand, the angular correlations of the temperature anisotropies in the CMB (Cosmic Microwave Background) predicted by cosmological inflation (in a rather generic, i.e., model-independent way) agree with very good precision with the anisotropies observed by WMAP \cite{WMAP1}, \cite{WMAP2} and Planck \cite{Planck}.

If inflation is defined as a phase of accelerated expansion of the universe, i.e., the scale factor $a(t)$ of the universe obeys $\ddot a >0$ (where a dot denotes a derivative w.r.t. cosmological time $t$), then it follows as an immediate consequence of the cosmological evolution equations (the FRW (Friedman-Robertson-Walker) equations) that the "matter" driving inflation must have negative pressure during the inflationary phase. The simplest way to introduce negative pressure into the FRW equations is by coupling a standard scalar field $\phi$ (the "inflaton") to gravity. This will provide negative pressure as long as its potential energy dominates over the kinetic energy \cite{linde2}. And, indeed, standard single scalar field models of inflation are ubiquitous in the literature on cosmological inflation (see, e.g., \cite{enzy}), and some of them reproduce the known cosmological observables very well. Some generalisations consist in allowing for a non-standard kinetic term for the inflaton field - the so-called k-inflation \cite{k-infl} - or in introducing several scalar fields (multi-field inflation) - see, e.g., \cite{wands} for some of the huge number of resulting possible scenarios. Unfortunately, already in the simplest case of a standard single scalar field model - consisting of the standard kinetic term and a potential - the available cosmological observations severely underdetermine the model (the potential, which a priory may be assumed to be a rather arbitrary function of the scalar field). This is related to the fact that currently the detailed microscopic mechanism responsible for inflation is not known, and the character of the inflaton field (e.g., as a fundamental or an effective field) is presently undetermined. In this paper, we shall take an agnostic point of view about these issues and treat scalar field inflation essentially as a useful {\em parametrisation} of inflation, where both a more profound understanding of its microscopic origin and a more precise determination of viable models require additional theoretical and observational progress.

Scalar field inflation is well-known to have a first integral relating the Hubble expansion parameter and the scalar field potential by a first-order differential equation which bears some similarity with the Hamilton-Jacobi equation of classical mechanics. The methods to derive this expression are known under the names of "Hamilton-Jacobi approach" \cite{bond1990}, \cite{kinney1997}, "fake supersymmetry" (or "fake supergravity") \cite{nunez2004}, \cite{skenderis2006}, the "superpotential method" \cite{garriga2016}, or  the "first-order formalism" \cite{bazeia1}, \cite{bazeia3}, \cite{FOEL}.
The resulting "Hamilton-Jacobi equation" (or "superpotential equation") permits to find exact inflationary solutions by simply choosing an exact expression $W(\phi)$ for the superpotential $W$, for an overview see, e.g., \cite{cher}. There exist related methods to generate exact solutions, based, e.g., on the choice of an exact expression for $\phi (t)$ \cite{Barrow1}, \cite{Barrow2}. Simple expressions for $W(\phi)$, however, usually lead to rather complicated inflaton potentials $V(\phi)$. From a field theory perspective, it is plausible to assume that the inflaton potential $V$ is more fundamental (the superpotential being just a useful calculational device), therefore the physical relevance of these exact solutions is questionable. 
Further, in \cite{kir1}, \cite{kir2} the superpotential equation was used to derive a kind of "renormalisation group equation" which permits to study different universality classes of inflationary models in a rather model-independent fashion (see our footnote below Eq. (\ref{dN})).
The most relevant point for our objectives is that the superpotential equation provides an inflationary evolution equation which is completely equivalent to the standard FRW evolution. Nevertheless,  inflationary cosmological evolution is usually studied using the original FRW equations.

It is the main purpose of the present paper to exploit the simplicity of the superpotential equation by putting it at the center stage of inflationary calculations. We will find that {\em all} observables of scalar field inflation can be calculated directly and in a simple fashion from the superpotential $W$ (the solution of the superpotential equation). In particular, 

i) the superpotential equation allows for a simple and direct calculation of the slow-roll expansion to arbitrary order. The slow-roll expansion is, in fact, strictly equivalent to a formal power series expansion of the superpotential equation; 

ii) the superpotential equation is a simple, mildly nonlinear first-order equation, and can, therefore,  easily be integrated numerically, e.g., with the help of a simple Mathematica program, which allows, in turn, to calculate all inflationary observables;

iii) in many cases, a Taylor series expansion of the superpotential may be performed, providing a reliable estimate for the radius of convergence of the slow-roll expansion;

iv) in many cases, the superpotential method allows for an exact albeit numerical determination of the exact slow-roll attractor (the particular slow-roll inflationary solution to which generic slow-roll inflationary solutions converge).

As remarked above, some single-field inflationary models reproduce the known observations very well, and the data even seem to prefer single-field inflation. We shall, therefore restrict our considerations to the standard single-field case. It should be mentioned, however, that the superpotential method may be easily generalised to k-inflation (where the superpotential equation continues to be a first-order ODE (ordinary differential equation), but with a stronger nonlinearity) or to the multi-field case, where the superpotential equation turns into a first-order PDE (partial differential equation).

We use natural units such that $c=\hbar =1$ throughout this paper. The only dimensionful parameter in the theory is, therefore, the reduced Planck mass $M_{\rm P}$ related to Newton's constant via 
\be
M_{\rm P} = \frac{1}{\sqrt{8\pi G}} \equiv \frac{1}{\sqrt{2\kappa}}
\ee
where we shall frequently use $\kappa = 4\pi G$. 

Our paper is organized as follows. In Section II, we briefly review the standard FRW cosmological evolution and introduce the superpotential. In Section III, we introduce the slow-roll parameters and demonstrate that the slow-roll expansion of the exact parameters may be achieved in a simple fashion by a formal power series expansion of the superpotential equation. To make our exposition self-contained,  in Section IV we briefly review the calculation of inflationary observables as well as the most recent observational bounds. In Section V, we review some general properties of the superpotential equation. Then, we solve the superpotential equation and calculate the corresponding inflationary observables for several specific models of single-field inflation. Finally, Section VI contains our summary.

\section{The FRW equations and the superpotential}
Cosmology is supposed to describe the universe at the largest scales, which justifies to assume its homogeneity and isotropy in a first approximation (inhomogeneities are then introduced as small perturbations). Further, observations strongly indicate that the universe is spatially flat (i.e., hypersurfaces of constant cosmological time have zero spatial curvature). With these assumptions, the metric relevant for cosmological evolution (the FRW metric) may be expressed like
\be
ds^2  = dt^2 - a^2 (t) \delta_{ij} dx^i dx^j .
\ee
Here, $t$ is called cosmological time, $a(t)$ is the scale factor describing isotropic expansion, and $(x^1, x^2 ,x^3 )$ are comoving cartesian coordinates -  distances between points corresponding to fixed values of the comoving coordinates expand with the general expansion described by $a(t)$. The Einstein equations may be derived from the variational problem for the action
\be \label{action}
\mathcal{S} = \mathcal{S}_{\rm EH} + \mathcal{S}_{\rm I}=-\frac{1}{4\kappa}\int d^4x\sqrt{|g|}\mathcal{R}
+ \int d^4 x \sqrt{|g|} \mathcal{L}_{\rm I}
\end{equation}
where $\mathcal{S}_{\rm EH}$ is the Einstein-Hilbert action (the minus sign results from our sign convention for the metric) and $\mathcal{S}_{\rm I}$ is the action for the inflaton field with the standard lagrangian density
\be
\mathcal{L}_{\rm I} = \frac{1}{2} g^{\mu\nu} \phi_{,\mu} \phi_{,\nu} - V(\phi) ,
\ee
where $V$ is a non-negative potential $V\ge 0$.
The Einstein equations are
\begin{equation}
G_{\mu\nu}=2\kappa T_{\mu\nu}
\label{Einsteineq}
\end{equation}
where $G_{\mu\nu}$ is the Einstein tensor, and the energy-momentum tensor is, in our case
\begin{equation}
T_{\mu\nu}(x)=2\frac{1}{\sqrt{|g|}} \frac{\delta}{\delta g^{\mu\nu}(x)} \mathcal{S}_{\rm I} = \phi_{,\mu}\phi_{,\nu}-g_{\mu\nu}\mathcal{L}_{\rm I} .
\label{Tgen}
\end{equation} 
If we insert the FRW metric into the Einstein equations, we have to assume a homogeneous inflaton field $\phi = \phi (t)$, for consistency. The energy-momentum tensor then simplifies to a perfect fluid form, 
\be
T_{00} = \frac{1}{2} \dot\phi^2 + V, \quad T_{ii} =  a(t)^2 \left( \frac{1}{2} \dot\phi^2 -V \right)
\ee
(and $T_{\mu\nu} =0$ for $\mu \not= \nu$),
or $T^\mu{}_\nu = {\rm diag} (\rho ,-p,-p,-p)$ where the proper energy density $\rho$ and pressure $p$ are given by
\be
\rho = \frac{1}{2} \dot\phi^2 + V , \quad p = \frac{1}{2} \dot\phi^2 - V.
\ee
The non-zero components of the Einstein tensor are (here, $a(t) =e^{A(t)}$)
\begin{equation}
G_{00}=3\dot{A}(t)^{2}, \quad 
G_{ii}=-e^{2A(t)}(3\dot{A}(t)^{2}+2\ddot{A}(t)) .
\label{Gii}
\end{equation}
Introducing now the Hubble function $H(t) \equiv \dot A = (\dot a/a)$, we get the two independent Einstein equations
\be \label{Einst1}
3H^2 = 2\kappa \left( \frac{1}{2} \dot \phi^2 + V \right) = 2\kappa \rho
\ee
and, after inserting this expression into $G_{ii}$, 
\be \label{Einst2}
\dot H = - \kappa \dot \phi^2 = -\kappa (\rho + p).
\ee
Observe that 
\be \label{ddot-a}
\frac{\ddot a}{a} = \dot H + H^2 = -\frac{\kappa}{3} (\rho + 3p),
\ee
so an accelerated expansion $\ddot a >0$ requires $\rho + 3p <0$, i.e., $\dot\phi^2 < V$ (the potential energy dominates over the kinetic one).

By varying the action w.r.t. the inflaton field, the field equation
\be \label{ddot-phi}
\ddot \phi + 3H\dot \phi + V_{,\phi} =0
\ee
may be derived, but this equation is not independent of the Einstein equations (it follows directly from the covariant conservation equation $\nabla_\mu T^\mu{}_\nu =0$, which for a general perfect fluid in the FRW metric reads $\dot \rho + 3 H (\rho + p)=0$). Equation (\ref{ddot-phi}) is formally equivalent to the classical equation of motion of a particle with coordinate $\phi$ under the influence of a force $F = - V_{,\phi}$ and a friction term $3H\dot \phi$.

Up to now, we assumed that all functions depend on cosmological time $t$. There exist, however, other possible choices for the time variable (the independent variable in our evolution equations) like, e.g., conformal time. In particular, in the superpotential formalism it turns out to be useful to consider the inflaton field itself as the new "time" variable during inflation.  Concretely, let us consider the Hubble function $H$ as a function of $\phi$ instead of $t$, i.e., 
\be \label{H-W}
H(t) = \frac{1}{\sqrt{\kappa}} W(\phi)
\ee
where $W(\phi)$ is the superpotential and the factor $1/\sqrt{\kappa}$ was introduced for convenience (such that $W$ is dimensionless). Inserting this expression into the second Einstein equation (\ref{Einst2}) we get $\dot H = (1/\sqrt{\kappa}) W_{,\phi} \dot \phi = -\kappa \dot \phi^2$, that is,
\be \label{phidot-W}
\dot \phi = - \kappa^{-\frac{3}{2}}W_{,\phi}.
\ee
As long as $\phi (t)$ is strictly monotonous ($\dot \phi$ is nonzero), it follows that $\phi$ is as good a time variable as $t$.
For a given superpotenial $W(\phi)$, the solution
\be
t = -\kappa^\frac{3}{2}\int_{\phi_0}^\phi \frac{d\tilde \phi}{W'(\tilde \phi)}
\ee
permits, in principle, to recover the cosmological time $t$ from the "inflaton time" $\phi$, but this step is never required for the calculation of inflationary observables. If we now insert (\ref{H-W}) and (\ref{phidot-W}) into the first Einstein equation (\ref{Einst1}), we get the
superpotential equation (see, e.g., \cite{kir1}, \cite{kir2})
\be
V = \frac{1}{2\kappa^2} \left( 3 W^2 - \frac{1}{\kappa} W_{,\phi}^2 \right)
\ee
which will play a central role in what follows. It is useful to rewrite this equation entirely in terms of dimensionless quantities.
Defining a dimensionless field and potential via 
\be \label{dim-less}
\phi = \frac{1}{\sqrt{\kappa}} \varphi , \quad U (\varphi)= 2\kappa^2 V(\frac{\varphi}{\sqrt{\kappa}} ) ,\quad \mathcal{W}(\varphi) =  W(\frac{\varphi}{\sqrt{\kappa}})
\ee
the dimensionless version of the superpotential equation is
\be
U(\varphi) = 3\mathcal{W}^2(\varphi )-  \mathcal{W}^2_{,\varphi}.
\ee
We shall find that the (exact, approximate or numerical) solution of this rather simple first-order equation is all that is required to calculate all inflationary observables we might be interested in for a given model (a given potential $U$).

\section{Slow-roll parameters}
\subsection{The Hubble and potential slow-roll parameters}
Cosmological inflation is defined as a phase of accelerated expansion, $\ddot a (t)>0$. With the help of eq. (\ref{ddot-a}) we may define a dimensionless parameter $\epsilon$ (the "first Hubble slow-roll parameter"),
\be
\frac{\ddot a}{a} = H^2 \left( 1 +\frac{\dot H}{H^2}  \right) \equiv H^2 (1-\epsilon )
\ee
such that accelerated expansion occurs for $\epsilon <1$. $\epsilon =0$ ($H=$ const.) corresponds to exponential expansion $a \sim e^{Ht}$ (de Sitter space-time), whereas $\epsilon <0$ is not possible for scalar field matter (it corresponds to "phantom matter" defined by $\rho + p <0$). Useful expressions for $\epsilon$ are
\be
\epsilon \equiv \epsilon_1 = - \frac{\dot H}{H^2} = \frac{W_{,\phi}^2}{\kappa W^2} = \frac{\mathcal{W}_{,\varphi}^2}{\mathcal{W}^2}.
\ee
For inflation to be of cosmological relevance, it is not enough that it occurs at a given instant, i.e., that $\epsilon <<1$ at a given time $t=t_0$. Inflation must also last sufficiently long to produce the required total expansion (a sufficiently large "number of e-folds"). This means that the "rate of change" of $\epsilon$ should be sufficiently small during inflation. A dimensionless expression imposing this condition is
\be
\epsilon_2 \equiv \frac{\dot \epsilon}{H\epsilon} <<1.
\ee
Using $\epsilon =-(\dot H/H^2)$ we easily calculate
\be
\epsilon_2 = \frac{\ddot H}{H\dot H} - 2 \frac{\dot H}{H^2} \equiv 2(\delta + \epsilon)
\ee
where we defined a second Hubble slow-roll parameter
\be
\delta = \frac{\ddot H}{2H\dot H} = - \frac{W_{,\phi\phi}}{\kappa W} = -\frac{\mathcal{W}_ {,\varphi\varphi}}{\mathcal{W}}
\ee
and used
\be
\dot H = - \kappa^{-2} W_{,\phi}^2 \; ,\quad \ddot H = 2\kappa^{-\frac{7}{2}}W_{,\phi\phi}W_{,\phi}^2 .
\ee
The second slow-roll condition now is $|\delta|<<1$ because $\delta$ does not have a definite sign, in general. 

Up to now, the relations defining the slow-roll parameters have been exact (they are called Hubble slow-roll parameters because they may be derived in terms of the Hubble function $H$). The disadvantage of this approach is that, in principle, we have to know the solution $H(t)$ (or $W(\phi)$) before we can decide about the smallness of $\epsilon$ and $\delta$. But the smallness of $\epsilon$ and $\delta$ implies that certain terms in the FRW equations are small and may be neglected in a first approximation. The resulting approximate FRW equations can be solved easily (essentially algebraically), and serve to find simplified expressions for $\epsilon$ and $\delta$, the so-called potential slow-roll parameters. We already know that $\epsilon <<1$ implies $\dot\phi^2 << V$. The intuitive picture is that the field $\phi (t)$ slowly rolls down the potential $V$ (explaining also the name "slow-roll inflation"). Neglecting $\dot\phi^2$ compared to $V$ in the first Einstein equation (\ref{Einst1}), we get the approximate equation
\be \label{H-aprox}
H^2\simeq \frac{2\kappa}{3}V.
\ee
To utilize the smallness of $\delta$, we use the second Einstein equation (\ref{Einst2}) to rewrite $\delta$ as
\be
\delta =  \frac{\ddot \phi }{ H\dot \phi }
\ee
 and conclude
\be
|\delta| <<1 \quad \Rightarrow \quad |\ddot \phi | << |H\dot \phi |  .
\ee
But in the field equation (\ref{ddot-phi}) this implies that the acceleration term $\ddot \phi$ can be neglected in comparison to the friction and force terms, leading to the approximate solution
\be \label{phi-aprox}
\dot \phi \simeq - \frac{V_{,\phi}}{3H} \quad \Rightarrow \quad \dot H \simeq -\kappa \frac{V_{,\phi}^2}{9H^2}.
\ee
Inserting the approximate expressions for $\dot H$ and $H^2$, we get for $\epsilon$
\be \label{eps-pot}
\epsilon =-\frac{\dot H}{H^2} \simeq \frac{V_{,\phi}^2}{4\kappa V^2} \equiv \epsilon_V
\ee
and, using in addition
\be
\ddot \phi \simeq - \frac{V_{,\phi\phi} \dot \phi}{3H} + \frac{V_{,\phi }\dot H}{3H^2} \simeq \frac{V_{,\phi\phi}V_{,\phi}}{9H^2} - \frac{\kappa V_{,\phi}^3}{27 H^4}
\ee
we get 
\be \label{del-pot}
\delta =  \frac{\ddot \phi}{H\dot \phi} \simeq -\frac{V_{,\phi\phi}}{2\kappa V} + \frac{V_{,\phi}^2}{4\kappa V^2} \equiv \delta_V + \epsilon_V
\ee
where $\epsilon_V$ and $\delta_V$ are called potential slow-roll parameters.
We remark that our definition of $\delta$ agrees with the one used in \cite{Weinberg}, but  $\delta = -\eta $ and $\delta_V = -\eta_V$ differ in sign from the slow-roll parameters $\eta$ and $\eta_V$  used in \cite{Baumann}. We also remark that, following standard practice, we are using two slightly different, although related, meanings of the term "slow-roll", which sometimes might lead to confusion. On the one hand, "slow-roll" refers to the existence of a region in field space where the Hubble slow-roll parameters are small. This "slow-roll" behavior is an intrinsic property of the particular model and the region in field space considered, giving rise to the most standard mechanism of single-field inflation, and is not related to any approximation. On the other hand, the term "slow-roll" also refers to a certain approximation. Indeed, Eqs. (\ref{H-aprox}) and (\ref{phi-aprox}) are the leading order slow-roll approximations to the Friedman and field equations, and the resulting expressions 
(\ref{eps-pot}) and (\ref{del-pot}) define the leading-order slow-roll approximations of the Hubble slow-roll parameters (i.e., the potential slow-roll parameters). Higher orders in this slow-roll approximation (i.e., the slow-roll expansion) can also be defined in a straight-forward manner, and we will discuss this issue within the superpotential framework in the next subsection.

Before doing so, let us introduce a more systematic method to obtain another set of useful slow-roll parameters,  the so-called Hubble-flow parameters \{$\epsilon_{n}$\} \cite{Schwarz} (see also \cite{Vennin}, \cite{Vennin2}).
For their definition, it is useful to introduce the number of e-folds $N$ as a new, dimensionless "time" variable,
\be
N(t) = \ln \frac{a(t)}{a(t_i)} = A(t) - A(t_i) 
\ee
(here $t_i$ is some reference  "initial" time, e.g, the time when inflation started). $N$ literally counts the 
number of exponential expansions which the universe (the spatial hypersurface) underwent. 
It follows that
\be \label{dN}
dN = Hdt = \frac{H}{\dot\phi} d\phi = -\kappa \frac{W}{W_{,\phi}} d\phi = - \frac{\mathcal{W}}{\mathcal{W}_{,\varphi}}d\varphi \equiv \pm \frac{1}{\sqrt{\epsilon}} d\varphi
\ee
showing the relation to other time variables. Here, $\sqrt{\epsilon}$ is the positive root, and the sign depends on whether $\varphi$ is monotonously growing or diminishing with 
$t$ (or with $N$)\footnote{
We remark that Eq. (\ref{dN}) may be re-expressed like $\kappa (d\phi/dN) = -(W_{,\phi}/W)$, which has the form of a renormalisation group (RG) equation. Indeed, if $\phi$ is interpreted as a "coupling constant", $N$ as the logarithm of a renormalisation scale, and $\beta (\phi) \equiv W_{,\phi}/W $ as a beta function, then the above equation is exactly equivalent to a RG equation \cite{kir1}, \cite{kir2}. Different choices for $\beta (\phi)$ then allow to study different "universality classes" of inflation without having to invoke specific models \cite{kir1}, \cite{kir2}, \cite{mabi}. Further, it was argued in \cite{kir1}, \cite{kir2} that this equivalence with a RG equation is not just formal but has a physical explanation in terms of the AdS/CFT correspondence.  
}.
In a slow-roll regime, space-time is close to de Sitter ($H$ is almost constant), so a plausible slow-roll expansion is in terms of the (small) deviations of $H$ from constancy. It is customary to consider the small variations of the Hubble distance $d_H = H^{-1}$, instead,  and to use the dimensionless time variable $N$.
Thus, let's define the dimensionless parameter 
\be
\epsilon_{0}=\frac{d_H}{d_{H_i}} = \frac{H_{i}}{H}
\ee
 (where $H_{i}$ is the Hubble parameter at $t_i$)  and the Hubble flow functions
\begin{equation}
\epsilon_{n+1}=\frac{d\ln\left|\epsilon_{n}\right|}{dN}.
\label{recurs}
\end{equation}
 $|\epsilon_{1}|<1 $ is required for inflation to occur, and the slow-roll regime corresponds to $\epsilon_n <<1 \;\; \forall \;\; n\ge 1$.

The first two slow-roll parameters coincide with the expressions derived above. Indeed, \begin{equation}
\epsilon_{1}=\frac{d\ln\epsilon_{0}}{dN}=- \frac{1}{H}\frac{d}{dt} \ln \frac{H}{H_i} =-\frac{\dot{H}}{H^{2}}
\label{EpT2}
\end{equation}
and
\begin{equation}
\epsilon_{2}=\frac{d\ln\epsilon_{1}}{dN}=\frac{\dot{\epsilon}}{\epsilon H}.
\label{Ep2T2}
\end{equation}
As before, simplified potential flow parameters $\epsilon_{n,V}$ may be obtained by using the simplified FRW equations (\ref{H-aprox}) and (\ref{phi-aprox}). Obviously,
\be
\epsilon_{0,V} = \sqrt{\frac{V_i}{V}}, \; , \quad \epsilon_{1,V} = \epsilon_V \; , \quad \epsilon_{2,V} = 2\delta_V + 4 \epsilon_V.
\ee
There exists, in fact, a systematic iterative expansion of the exact Hubble flow parameters $\epsilon_n$ in terms of higher-order potential flow parameters,
\be
\epsilon_n = \sum_{i=1}^\infty \epsilon_n^{(i)}
\ee
where $\epsilon_n^{(1)} \equiv \epsilon_{n,V}$ and the higher orders $i>1$ may be found by iteratively improving the simplified FRW equations  (\ref{H-aprox}) and (\ref{phi-aprox}). A detailed calculation of these higher $\epsilon_n^{(i)}$ can be found in \cite{Vennin}, \cite{Vennin2}. We shall not follow this approach here, because we will see in the next section that the superpotential equation allows for a very simple and systematic calculation of these higher-order corrections.

\subsection{The slow-roll expansion in the superpotential formalism}
For simplicity, we will use the dimensionless superpotential equation, which we rewrite as
\be \label{superpot-eq-beta}
U(\varphi) = 3\mathcal{W}^2(\varphi )- \beta \mathcal{W}^2_{,\varphi}.
\ee
Here, $\beta$ is a formal expansion parameter which may be set equal to $\beta =1$ at the end of the calculation. We introduce this parameter, because then the slow-roll expansion is just a formal power series expansion of the superpotential equation in $\beta$.
Indeed, rewriting
\be \label{W-exp}
\mathcal{W} = \sqrt{\frac{U}{3}} \sqrt{ 1 + \beta \frac{\mathcal{W}_{,\varphi}^2}{U}}
\ee
we may expand this expression in $\beta$. Expanding up to $\beta^0$ provides the leading slow-roll expansion, which we shall call $\mathcal{W}^{(1)}$ (so the expansion up to $\beta^n$ provides the slow roll approximation $\mathcal{W}^{(n+1)}$), that is $\mathcal{W}^{(1)} =\sqrt{U/3}$, leading to
\be \label{W1}
\quad \mathcal{W}_{,\varphi}^{(1)} = \frac{1}{\sqrt{12}} \frac{U_{,\varphi}}{\sqrt{U}},
\ee
so we get (remember $(H_i/H) = (\mathcal{W}_i / \mathcal{W})$)
\be
\epsilon_0^{(1)} = \sqrt{\frac{U_i}{U}}
\ee
and 
\be
 \epsilon_1^{(1)} = \left( \frac{\mathcal{W}_{,\varphi}^{(1)}}{\mathcal{W}^{(1)}}\right)^2 = \frac{1}{4}\frac{U_{,\varphi}^2}{U^2} = \epsilon_V.
\ee
In next-to-leading order we get
\be \label{W2}
\mathcal{W}^{(2)} = \sqrt{\frac{U}{3}}\sqrt{1+\frac{\beta}{12}\frac{U_{,\varphi}^2}{U^2}}
\simeq \sqrt{\frac{U}{3}} \left( 1 + \frac{\beta}{24}\frac{U_{,\varphi}^2}{U^2} \right)
\ee
and
\be
\mathcal{W}^{(2)}_{,\varphi} \simeq \sqrt{\frac{U}{3}}\, \frac{U_{,\varphi}}{2U} \left( 1 + \beta \left( \frac{1}{6}\frac{U_{,\varphi\varphi}}{U} - \frac{1}{8} \frac{U_{,\varphi}^2}{U^2} \right) \right)
\ee
leading to
\be
\epsilon_0^{(1)} + \epsilon_0^{(2)} = \epsilon_0^{(1)} \left( 1 - \frac{\beta}{6} \epsilon_1^{(1)} \right)
\ee
and to
\bea
\epsilon_1^{(1)} + \epsilon_1^{(2)} &=& \left( \frac{\mathcal{W}_{,\phi}^{(2)}}{\mathcal{W}^{(2)}}\right)^2 
\simeq \frac{1}{4}\frac{U_{,\varphi}^2}{U^2} \frac{\left( 1 + 2 \beta \left( \frac{1}{6}\frac{U_{,\varphi\varphi}}{U} - \frac{1}{8} \frac{U_{,\varphi}^2}{U^2} \right) \right) }{\left( 1 + 2\frac{\beta}{24}\frac{U_{,\varphi}^2}{U^2} \right)} \nonumber \\
&\simeq & \frac{1}{4}\frac{U_{,\varphi}^2}{U^2} \left( 1 + \beta \left( \frac{1}{3} \frac{U_{,\varphi\varphi}}{U} - \frac{1}{3}\frac{U_{,\varphi}^2}{U^2} \right) \right) 
= \epsilon_1^{(1)} \left( 1 - \beta \frac{\epsilon_2^{(1)}}{3} \right)
\eea
which for $\beta = 1$ exactly coincide with the next-to-leading slow-roll results of \cite{Vennin}.
In general, all higher order corrections may be obtained by i) using the recursion relation (\ref{recurs}) to obtain $\epsilon_n$ in terms of the superpotential, and ii) by expanding the superpotential up to the required order in $\beta$ (and only maintaining terms up to this order in the subsequent calculation).

\section{Inflationary observables}
One of the great successes of inflation is that it provides rather robust predictions for certain statistical properties of the small inhomogeneities which can be observed in the universe at sufficiently large scales. The two most important cases are the inhomogeneities in the matter distribution of the universe at sufficiently large scales (large scale structure (LSS)) and the temperature inhomogeneities in the CMB radiation. The general idea is that small fluctuations about the homogeneous inflaton and the FRW metric are unavoidably generated as a consequence of their quantum nature (essentially the uncertainty relation). These fluctuations are generated at microscopic scales but then are blown up to huge scales by inflation (for a detailed account of cosmological perturbations in inflation see, e.g., \cite{muk2}). A further important property is that, as a consequence of the inflationary evolution, these fluctuations evolve essentially unchanged once they reach a certain size ("horizon exit")\footnote{More precisely, the "power spectrum" characterizing these fluctuations becomes time-independent on superhorizon scales, that is, in the large wavelength limit $k|\tau| \sim (k/aH) <<1$. But once the functional form of this power spectrum in the large wavelength limit is determined, then, owing to its time independence, it may be evaluated at any instance $t$ (or conformal time $\tau$). It turns out useful to choose the "horizon exit time" $t_k$ for this evaluation, because then $t$ is no longer an independent variable but, instead, a function of the wave number $k$ defined implicitly by the horizon exit condition $k = a(t_k) H(t_k)$. This point is discussed in detail in the main text, see Section 4.A.}. This constant evolution continues until the post-inflationary ("standard big bang") universe catches up with the scale of the fluctuation ("horizon re-entry"), in which moment the inflationary fluctuations may serve as "seeds" for the inhomogeneities in the universe. In practical terms, this description only works on sufficiently large scales (sufficiently late horizon re-entry), because for smaller scales (earlier re-entry) the subsequent nonlinear evolution (gravitational clumping) essentially washes out all influences of inflationary fluctuations. In addition, small scales which re-enter too early are affected by the poorly known evolution of the post-inflationary universe before radiation domination
(i.e., in the so-called reheating phase, where the energy density stored in the inflaton field is converted into standard matter).

\subsection{Fluctuations}
More concretely, due to the symmetries of the FRW metric, small fluctuations about it and about the homogeneous inflaton field may be classified into (spatial) scalar, vector and tensor fluctuations. Vector fluctuations cannot be generated by inflation and, in addition, are suppressed quickly during inflationary evolution, so they may be ignored. There are two tensor modes corresponding to the two polarisations of the primordial gravitational waves which these two modes will induce. The most important fluctuations for cosmological predictions, however, are the scalar ones. General scalar fluctuation about the FRW metric may be described by four scalar functions. Not all of these fluctuations are physical,  because of the freedom to perform two scalar coordinate transformations. In cosmological perturbation theory, {\em gauge transformations} are frequently considered instead of coordinate transformations, where a gauge transformation is a coordinate transformation which acts only on the space-time manifold given by the full metric with the fluctuations included, $g_{\mu\nu} = \bar g_{\mu\nu} + \delta g_{\mu\nu}$, but not on the  background space-time given by the FRW metric $\bar g_{\mu\nu}$. The two scalar gauge functions, thus, allow to eliminate two scalar fluctuations such that, e.g., in Newtonian gauge, the scalar metric fluctuations may be expressed by two scalar fluctuation functions $\Phi$ and $\Psi$ as
\be
\delta g_{00} = 2\Phi ,\quad \delta g_{ij} = 2a^2 \delta_{ij} \Psi, \quad \delta g_{0i}=0.
\ee
The background energy-momentum tensor (EMT) $\bar T_{\mu\nu}$ inserted into the FRW equations is of the perfect fluid type, by construction, but this is not necessarily true for the fluctuation part. A completely general fluctuation EMT $\delta T_{\mu\nu}$ contains four scalar functions. If the total EMT $T_{\mu\nu} = \bar T_{\mu\nu} + \delta T_{\mu\nu}$ still describes a perfect fluid (as is the case, e.g., for single-field inflation), then one of these four scalar functions (related to dissipation) is zero. This implies that on-shell (i.e., when the linearized Einstein equations for the fluctuations are invoked), the two scalar metric functions are equal, $\Phi = \Psi$. Further, for single-field inflation, the three remaining scalar fluctuation functions of the EMT (the energy density fluctuation $\delta \rho$, pressure fluctuation $\delta p$ and scalar velocity potential fluctuation $\delta u$) may all be expressed in terms of the inflaton fluctuation $\delta \phi$ (and its time derivative) and the metric fluctuation $\Psi$. In other words, there are two relevant scalar fluctuations in the case of single-field inflation. It turns out, however, that there is a particular combination of the two,
\be \mathcal{R} \equiv -\Psi + H\delta u,
\ee
which is of special importance. First of all, the fluctuation $\mathcal{R}$ is gauge invariant (although its geometrical interpretation depends on the chosen gauge). This implies that the perturbation $\mathcal{R }$ may be consistently inserted into the action on its own (i.e., gauge transformations do not turn on further fluctuations). In the limit of small fluctuations, one then restricts to a fluctuation action quadratic in $\mathcal{R}(t,\vec x)$. In a next step, the Fourier transform $\mathcal{R}(t,\vec x)= \int (d^3 k/(2\pi)^{3/2}) \mathcal{R}(k,t)\exp (i\vec k\cdot \vec x)$ is inserted.  Here, $\vec x$ are comoving coordinates, so $\vec k$ is the comoving wave vector and $k\equiv |\vec k|$ the comoving wave number. Further, the Fourier transform $\mathcal{R}(k,t)$ only depends on the wave number $k$ (and not on $\vec k$) as a result of the isotropy of the background. In the linear approximation (quadratic action), different modes $\mathcal{R}(k,t)$ do not interact and evolve independently, and we are left with a sum (integral) of independent one-dimensional mechanical systems labeled by the wave number $k$. 

This mechanical system may be brought into the form of a correctly normalized standard harmonic oscillator with a time-dependent frequency by a transformation which acts both on the independent and the dependent variables. For the independent variable, cosmological time $t$ is replaced by conformal time
\be \tau = \int_{t_e}^t \frac{dt'}{a(t')}.
\ee
Here, the choice of the integration constant $t_e$ is not relevant for the transformation, but it will be relevant for some physical considerations, later on. Concretely, we assume that $t_e$ is the time when inflation ends, defined by $\epsilon (t_e)=1$, which implies that the coordinate $\tau$ is negative during inflation, whereas the post-inflationary (standard big bang) evolution starts at $\tau =0$. The reason for this choice is as follows. During slow-roll inflation the scale factor $a(t)$ may be approximated reasonably well by the exponential de Sitter evolution $a(t) = a_* \exp (\bar Ht)$ with a constant Hubble parameter $\bar H$ (here $a_*$ is a constant which is irrelevant for our purposes).  If we assume that this approximation is good sufficiently long, i.e.,  until the instant $t_s$ where $t<<t_s<t_e$, then $|\tau|$ during inflation may be approximated by the co-moving Hubble distance (co-moving curvature radius; we use $\bar H = H(t)$)
\be
|\tau | =   \frac{1}{H(t)} \left( \frac{1}{a(t)} - \frac{1}{a(t_s)} \right) + \int_{t_s}^{t_e} \frac{dt'}{a(t')}
\le \frac{1}{a(t)H(t)} \left( 1  - \frac{a(t)}{a(t_s)}  + H(t)(t_e - t_s)\frac{a(t)}{a(t_s)} \right)
\ee
that is
\be
 |\tau | \sim \frac{1}{H(t)a(t)}
 \ee
 because the two remaining terms are suppressed by a factor $a(t)/a(t_s)$.\footnote{As a matter of fact, even for the post-inflationary evolution $t>t_e$ the comoving event horizon $\tau (t) $ is frequently approximated by the comoving Hubble distance $(H(t)a(t))^{-1}$. E.g. for the current age of the universe $t_0$, the event horizon is about $16$ Gly (= Giga light years) and the Hubble distance about $14 $ Gly, so the approximation is not too bad.} 
 
 The dependent variable is transformed like $v(k,t) = z\mathcal{R}(k,t)$, $z^2 \equiv a^2 \dot \phi^2/H^2 = (1/\kappa) a^2 \epsilon$, where $v$ is called the Mukhanov variable.  The action in terms of these new variables gives rise to the so-called Mukhanov-Sasaki equation
 \be \label{MuSaEq}
 v'' + \left(k^2 -\frac{z''}{z}\right) v =0
 \ee
 (here $' \equiv d/d\tau$),
 which is just the equation of motion of a harmonic oscillator with time-dependent frequency. The Mukhanov variable $v$ is now promoted to a quantum operator, $v(k,\tau) \to {\bf v}_{\vec k} \equiv v_k (\tau) {\bf a}_{\vec k} + v^*_k(\tau) {\bf a}^\dagger_{-\vec k}$, where the correct choice of the mode function $v_k (\tau)$ (a certain solution of eq. (\ref{MuSaEq})) may be found from the observation that in the limit $|k\tau| >>1$ equation (\ref{MuSaEq}) approaches the standard flat space harmonic oscillator $v'' + k^2 v =0 $. The idea of this quantization is that, as a consequence of its quantum nature, the field ${\bf v}_k$ unavoidably generates quantum fluctuations all the time. These fluctuations are generated at tiny scales (this is somewhat hidden in the expressions, due to our convention $\hbar =1$), but the subsequent expansion magnifies these scales to macroscopic sizes. Indeed, while the comoving scale $\lambdabar_k = k^{-1}$ of a fluctuation is fixed, its physical size $d_k = a(t) \lambdabar_k$ grows with $a(t)$.  
 
Obviously, the quantum fluctuations of ${\bf v}$ cannot be used to predict the matter density or temperature fluctuations in the universe, because the latter are random. It is, however, possible to relate the mean square expectation value $\langle {\bf v}_{\vec k} {\bf v}_{\vec k'} \rangle$ to the (classical statistical) auto-correlation functions (two-point functions) of density or temperature fluctuations.  For the correctly normalised harmonic oscillator (and for our choice of conventions for the Fourier transform), the creation and annihilation operators obey the commutator
$[{\bf a}_{\vec k},{\bf a}^\dagger_{\vec k'}] = \delta^3 (\vec k - \vec k')$, leading to the mean-square (quadratic) fluctuations
\be
\langle {\bf v}_{\vec k} {\bf v}_{\vec k'} \rangle = |v_k(\tau)|^2  \delta^3 (\vec k - \vec k') \equiv \mathcal{P}_v (k,\tau) \delta^3 (\vec k - \vec k')
\ee
and to (here the operator $\widehat{\mathcal{R}}(\vec k, \tau)$ is related to $\mathcal{R}(k,\tau)$ like ${\bf v}_{\vec k}$ is related to $v(k,\tau)$)
\be
\langle \widehat{\mathcal{R}}(\vec k,\tau )\widehat{\mathcal{R}}(\vec k' ,\tau )\rangle = |\mathcal{R}(k,\tau)|^2 \delta^3 (\vec k - \vec k') \equiv \mathcal{P}_\mathcal{R} (k,\tau) \delta^3 (\vec k - \vec k')
\quad \Rightarrow \quad \mathcal{P}_\mathcal{R} = z^{-2} \mathcal{P}_v .
\ee
$\mathcal{P}_\mathcal{R}(k,\tau)$ has the important property that it becomes time-independent (constant) in the limit of late time (large wave length) $k|\tau| \to 0$. This is true under rather general circumstances (see, e.g., \cite{Weinberg}), and this fact is important because it implies that the power spectrum $\mathcal{P}_\mathcal{R}(k,\tau)$ for sufficiently large wavelengths remains unaltered also during the reheating phase. On the other hand, the demonstration of this constancy is much easier for the slow-roll inflation phase. Indeed, in the slow-roll approximation, the relevant mode function $v(k,\tau)$ behaves in the limit $k|\tau| \to 0$ like (see appendix A)
\be \label{nu}
v^0(k,\tau) \equiv \lim_{k|\tau| \to 0} v(k,\tau) = c \, 2^{\nu -2}k^{-\nu} |\tau|^{\frac{1}{2} -\nu} \; , \quad \nu \equiv \frac{3}{2} + 2\epsilon +\delta ,
\ee
($c$ is an irrelevant constant phase)
and this $\tau$ dependence is exactly cancelled by the $\tau$ dependence of $z^{-1}$ (remember $\mathcal{R} = z^{-1} v$), i.e., $z \sim |\tau|^{\frac{1}{2} -\nu}$, as we show now. Indeed, calculating the log derivative (and using $(d/d\tau) = a (d/dt)$), we get
\be
\frac{\tau}{z}\frac{dz}{d\tau} = \tau \left(\dot a + \frac{a\dot \epsilon}{2\epsilon}\right) =
\tau aH \left(1+\frac{1}{2}\epsilon_2 \right) .
\ee
Here, $\epsilon_2$ is considered constant in the slow-roll approximation. For the $\tau$ dependence of $aH$ we get
\be
\frac{d}{d\tau}\frac{1}{aH} = a\frac{d}{d t}\frac{1}{aH} = -\frac{1}{aH^2}\left( \dot a H + a \dot H \right) = -1+\epsilon
\ee 
which is constant within slow roll, and, therefore,  
\be
aH \simeq \frac{-1}{(1-\epsilon) \tau}.
\ee
Altogether, we find
\be
\frac{\tau}{z}\frac{dz}{d\tau} \sim - (1+\frac{1}{2} \epsilon_2 + \epsilon ) =\frac{1}{2} -\nu
\ee
as announced, demonstrating the constancy of $\mathcal{R}^0(k)\equiv \lim_{k|\tau| \to 0} \mathcal{R}(k,\tau)$. We now know the functional (in)dependence of $\mathcal{R}^0$, but we still need its absolute value or, equivalently, the value of 
\be \label{PR0}
\mathcal{P}_{\mathcal{R}^0}(k) = z^{-2}|v^0|^2 =2^{2\nu -4} \frac{\kappa}{k^3} k^{3-2\nu}\frac{|\tau|^{1-2\nu}}{a^2\epsilon}
\ee 
where we split off a canonical $k^{-3}$ dependence which follows on purely dimensional grounds. $\mathcal{P}_{\mathcal{R}^0}$ does not depend on time, therefore we may choose any value of $t$ (or $\tau$) for its explicit evaluation. It is customary to choose the instant when the reduced comoving wave length $\lambdabar_k$ is equal to the comoving Hubble radius (curvature radius), i.e., $\lambdabar_k = (a(t_k)H(t_k))^{-1}$ or
\be
k = a(t_k)H(t_k).
\ee
We emphasize that $t_k$ here is {\em not} an independent variable, but a function of $k$ defined implicitly by this equation. The reason for this choice is that now everything in Eq. (\ref{PR0}) explicitly only depends on $k$ and not on $\tau$, which simplifies the comparison with observable quantities (of course, (\ref{PR0}) really only depends on $k$ in any case, but this is not true for the individual factors appearing in  (\ref{PR0}) for other choices of $t$, which would complicate the evaluation). Using $t_k$ and $|\tau_k| \simeq ((1-\epsilon)aH)^{-1}\vert_{t_k}$ we finally get
\be
\mathcal{P}_{\mathcal{R}^0}(k) =\left. 2^{2\nu -4} \frac{\kappa}{k^3} \frac{(aH)^{3-2\nu}}{((1-\epsilon)aH)^{1-2\nu}}\frac{1}{a^2\epsilon} \right|_{t_k}= \left.  \frac{2^{2\nu -4}\kappa}{(1-\epsilon)^{1-2\nu}k^3}\frac{H^2}{\epsilon}\right|_{t_k} \simeq \frac{\kappa}{2k^3} \left. 
\frac{H^2}{\epsilon}\right|_{t_k}
\ee
where expressions like $1-\epsilon$ were replaced by $1$ in the last step. 

\subsection{Observables}
To compare with observations, it is useful to define certain dimensionless quantities.
The dimensionless scalar power spectrum $\Delta^2_s$ is
\be \label{Del-s}
\Delta^2_s (k) \equiv \frac{k^3}{2\pi^2} \mathcal{P}_{\mathcal{R}^0}(k) = \left. \frac{\kappa}{4\pi^2}\frac{H^2}{\epsilon}\right|_{t_k}.
\ee
If $H$ and $\epsilon$ depend on time, then $\Delta^2_s$ will depend on $k$ (will not be scale invariant). This is usually quantified by the scalar spectral index $n_s$, defined as the log derivative
\be
n_s -1 \equiv \frac{d \ln \Delta^2_s}{d \ln k}.
\ee
It is useful to replace the derivative w.r.t. $\ln k$ by the derivative w.r.t. the e-fold number $N$, because then everything can be expressed directly in terms of the Hubble flow parameters $\epsilon_n$. With $k = aH\vert_{t_k} \; \Rightarrow \; \ln k = N + \ln H$ we get within slow roll
\be
\frac{d \ln k}{dN} = 1 - \epsilon_1 \simeq 1
\ee
and (evaluation at $t=t_k$ is understood)
\be
n_s -1 \simeq \frac{d}{dN} \left( 2\ln H - \ln \epsilon \right) = -2\epsilon_1 - \epsilon_2 = -4\epsilon - 2 \delta .
\ee

The tensor fluctuation modes which give rise to the seeds of primordial gravitational waves can be treated in a similar fashion. The tensor fluctuations are $\delta g_{ij} = a^2 h_{ij}$ where $h_{ij}$ is a traceless, transverse tensor which describes the two polarization modes of gravitational waves. The quadratic fluctuation action may be calculated and the correctly normalized mode functions  may be found and then quantized, like in the scalar case. Both polarizations give the same contribution to the power spectrum, and the resulting dimensionless tensor power spectrum is
\be \label{Del-t}
\Delta^2_t (k) = \frac{4\kappa}{\pi^2} \left. H^2 \right|_{t_k}.
\ee
Finally, the tensor spectral index is 
\be
n_t \equiv \frac{d\ln \Delta^2_t}{d\ln k} \simeq -2\epsilon_1 = -2\epsilon.
\ee
We can also define the tensor-to-scalar ratio
\be
r = \frac{\Delta^2_t (k)}{\Delta^2_s (k)}= 16 \epsilon .
\ee

The latest Planck measurements \cite{Planck} lead to the following observed values or constraints for the inflationary observables. First of all, for $n_s$ the following best fit value is found,
\be
n_s = 0.965 \pm 0.004 .
\ee
No variation of $n_s$ with $k$ is found in \cite{Planck}, so this quantity seems to be constant in our universe. The scalar power spectrum is evaluated at the pivot scale $k_* = 0.002 \, ({\rm Mpc})^{-1}$, i.e., $1/k_* = 500 \, {\rm Mpc}$ (Mpc = Megaparsec). This is about $1/8$ of the Hubble distance of the current universe, $1/H_0 \simeq 14 \, {\rm Gly} \simeq 4 \, {\rm Gpc}$.  The value quoted in \cite{Planck} is
\be
\Delta^2_s (k_*) = (2.099 \pm 0.028) \cdot 10^{-9} \; , \quad \Delta^2_s (k) = \left(\frac{k}{k_*}\right)^{n_s -1} \Delta^2_s (k_*) .
\ee
The small value of $n_s -1$ implies that the running with $k$ is very slow.
Tensor fluctuations can, in principle, be measured not only from gravitational waves but also from polarisations of the CMB, but the results are, up to now, inconclusive. This leads to an upper bound for
the tensor-to-scalar ratio $r$ \cite{Planck}
\be
r \le 0.064
\ee
(here we assume that the constancy of $n_s$ is inherited by $\epsilon$ and, therefore, by $r$ within single-field inflation). Finally, no nontrivial restriction on $n_t$ has been found. 

The observed values and constraints on $n_s$ and $r$ directly translate into constraints on the slow-roll parameters $\epsilon$ and $\delta$. The value for $\Delta_s^2$, on the other hand, implies a condition on the absolute value (coupling strength) of the inflaton potential, because in lowest order in the slow-roll expansion we have
\be
\left( \Delta_s^2 \right)_V = \frac{2\kappa^3}{3\pi^2} \frac{V^3}{V_{,\phi}^2}= \frac{1}{3\pi^2} \frac{U^3}{U_{,\varphi}^2}= \frac{1}{12 \pi^2}\frac{U}{\epsilon_V}.
\ee  
This constraint may always be satisfied by choosing the right value for the overall strength of the inflaton potential, without influencing the remaining constraints. We will, therefore, not further consider this constraint, because it only provides nontrivial information once the overall strength of the inflaton potential is a prediction of theory, which we do not assume here.

For concrete evaluations, we still have to identify the relevant scales $1/k$ which may be related to observations in the post-inflationary universe. 
A co-moving scale $1/k$ which crossed the horizon during inflation at a time $t_k^<$ such that $k=a(t_k)H(t_k) \equiv a_k H_k$, will remain outside the horizon during the rest of inflation, because the co-moving horizon $(aH)^{-1}$ shrinks during inflation while $1/k$ is constant. It will re-enter the horizon (and start to influence the physics of the universe) during the post-inflationary FRW evolution, at an instant $t_k^>$ such that $a(t_k^>) H(t_k^>) = a_kH_k =k$. In other words, if we want to know when a mode which entered the horizon at $t_k^>$ during FRW expansion exited the horizon during inflation, we have to match the  inflationary and post-inflationary evolutions of the co-moving Hubble distance $(aH)^{-1}$. It is customary to use the required number of e-folds (here $a_e \equiv a(t_e)$ where inflation ends at $t_e$)
\be
N_k = \ln \frac{a_e}{a_k}
\ee
for the description of the inflationary part. We start from the identity
\be
\frac{k}{a_0H_0} = \frac{a_kH_k}{a_0H_0} = \frac{a_k}{a_e}\frac{H_k}{H_e} \frac{a_eH_e}{a_0H_0}
\ee
where $(a_0H_0)^{-1}\equiv (a(t_0)H(t_0))^{-1}$ is the co-moving Hubble distance of the current universe (and $t_0$ its age). This allows to express $N_k$ like
\be
N_k = - \ln\frac{k}{a_0H_0} + \ln \frac{H_k}{H_e} + \ln \frac{a_eH_e}{a_0H_0}.
\ee
This identity is converted into a useful expression, allowing to match inflationary and post-inflationary evolution, by assuming that the factor $H_k/H_e$ is described by inflation (using slow-roll), whereas $(a_eH_e)/(a_0H_0)$ is described by the post-inflationary FRW evolution \cite{li-le}. The phases of the FRW evolution corresponding to radiation, matter and dark energy domination are known and may be calculated explicitly. The evolution during the reheating phase is not known, but may be estimated. The resulting expression used, e.g., in \cite{Planck}, is 
\be
N_k \simeq 67 - \ln\frac{k}{a_0H_0} + \frac{1}{4}\ln \frac{V_k}{M_{\rm P}^4} + \frac{1}{4}\ln \frac{V_k}{\rho_e}
+ \frac{1-3w_{in}}{12(1+w_{in})}\ln \frac{\rho_{th}}{\rho_e} - \frac{1}{12} \ln g_{th}.
\ee
Here, the first term on the r.h.s. (the number 67) contains all the known FRW results as well as some numerical factors split off the remaining logarithms. The second term just expresses that smaller scales $1/k$ need less e-folds. The third and fourth term come from the inflationary phase. The third term means that a smaller potential energy density $V_k = V(\phi(t_k))$ at horizon exit requires less e-folds. The fourth term reflects the (small) scale dependence of the Hubble parameter during inflation ($\rho_e$ is the energy density at $t_e$). The last two terms are estimated corrections from the reheating phase. Here, $\rho_{th}$ is the energy density at thermalization (the instant $t_{th}$ when the net conversion of inflaton energy density into matter energy density stops), and $g_{th}$ is the effective number of bosonic degrees of freedom at the same instant. Further, $w_{in} = p_{in}/\rho_{in}$ characterizes a model-dependent effective equation of state in the intermediate phase between $t_e$ and $t_{th}$.  In any case, the contributions of these last two terms are usually small \cite{Planck13}. The most relevant contribution in most models comes from the third term (it contributes about $-10$) and, depending on the scale $1/k$, from the second term [in cosmological applications, typically $(a_0H_0)^{-1} \ge (1/k) \ge 10^{-4} (a_0H_0)^{-1}$]. As a result, the range of $N_k$ usually considered for large scales (not much smaller than $(a_0H_0)^{-1}$)  is $50 \le N_k \le 60$, and we shall stick to this assumption.

\section{The superpotential method for particular potentials}
In this section, we will, in a first step, discuss some general properties of the superpotential equation. Then we calculate the superpotential and the resulting inflationary observables for some particular  potentials. Concretely, we shall consider the cases of exponential and quadratic potential and the cases of hill-top and Starobinsky inflation. The first two cases do not lead to realistic models of inflation, but they are simple and allow to discuss some qualitative features of the method. The cases of hill-top and Starobinsky inflation, on the other hand, are realistic models in the sense that they are compatible with all observational restrictions. 

\subsection{General properties of the superpotential}

Before starting the specific calculations, it is useful to review some general properties and implications of the superpotential equation (\ref{superpot-eq-beta}). First of all, it is an equation which is quadratic in $\mathcal{W}_{,\varphi}$, leading to the two roots (for the moment, we retain the formal expansion parameter $\beta$, which must be set to $\beta =1$ at the end)
\be
\mathcal{W}_{,\varphi} = \pm \sqrt{\beta^{-1} (3\mathcal{W}^2 - U)}.
\ee
For each model, only one of the two signs is acceptable. The correct choice follows from the condition $\dot H \le 0$, and from the identity $\dot H = 1/\sqrt{\kappa} \mathcal{W}_{,\varphi} \dot\varphi $, and reads
\be
\mathcal{W}_{,\varphi} = - {\rm sign}(\dot\varphi ) \sqrt{\beta^{-1} (3\mathcal{W}^2 - U)}.
\ee
So if $\varphi$ increases while rolling down the potential, then the negative sign must be chosen, and vice versa. For explicit calculations, it turns out to be useful to split off the potential $U$ from the superpotential equation. That is to say, we define a function $f$ via $\mathcal{W} = \sqrt{U}f$ and divide the superpotential equation (\ref{superpot-eq-beta}) by $U$ (remember that $U>0$ is required for inflation to occur), resulting in 
\be \label{f-eq}
1 = 3f^2 - \beta \left( \frac{1}{2}\frac{U_{,\varphi}}{U}f + f_{,\varphi}\right)^2 = 3f^2 - \beta \left[ {\rm sign} (U_{,\varphi}) \sqrt{\epsilon_V}f + f_{,\varphi}\right]^2 
\ee
with the relevant root
\be \label{f-eq-root}
f_{,\varphi} = - \frac{1}{2}\frac{U_{,\varphi}}{U}f - {\rm sign}(\dot\varphi ) \sqrt{\beta^{-1} (3f^2 - 1)}.
\ee
The function $f$ is always positive (because $H=(1/\sqrt{\kappa})\mathcal{W} >0$) and, therefore, obeys the inequality $f\ge (1/\sqrt{3})$. We may obtain more information by comparing the leading order slow-roll parameter $\epsilon_V =(U_{,\varphi}^2/(4 U^2))$ with the exact slow-roll parameter
\be \label{eps-f}
\epsilon = \frac{\mathcal{W}_{,\varphi}^2}{\mathcal{W}^2} = \frac{3\mathcal{W}^2 - U}{\mathcal{W}^2} = \frac{3f^2 -1}{f^2}.
\ee
Firstly, inflation occurs in the interval $0\le \epsilon \le 1$, that is $(1/\sqrt{3}) \le f \le (1/\sqrt{2})$. 
Secondly, it follows from Eq. (\ref{eps-f}) that $\epsilon$ is also bounded from above, $0\le \epsilon \le 3$, in contrast to $\epsilon_V$ which only obeys $\epsilon_V \ge 0$.  This bound can be useful to check the precision of numerical calculations. Eq. (\ref{eps-f}) also serves to estimate the reliability of the slow-roll approximation from a (exact, power-series-expansion or numerical) solution $f$. Slow-roll is reliable as long as $\epsilon$ is small, and $\epsilon$ is small if $f-(1/\sqrt{3})<<(1/\sqrt{3})$.  

\subsection{Initial conditions and attractor solutions}
Eq. (\ref{f-eq-root}) is a first-order ODE, therefore there exists a one-parameter family of solutions, and choosing a particular solution requires to impose one initial condition. For concreteness, we assume that $\dot\varphi >0$, i.e., $\varphi$ grows while it rolls down the potential, which implies $U_{,\varphi}<0$. Eq. (\ref{f-eq-root}) then reads (we also set $\beta \equiv 1$)
\be \label{f-root-1}
f_{,\varphi} = \left| \frac{1}{2}\frac{U_{,\varphi}}{U}\right| f -  \sqrt{ (3f^2 - 1)} = f\left( \sqrt{\epsilon_V} - \sqrt{\epsilon} \right)
\ee
(the discussion for $\dot \varphi <0$ is completely analogous and just requires some sign changes).

First, let us assume a generic potential $U$ which gives rise to slow-roll inflation, i.e.,  there exist field values $\varphi_i$ such that $\epsilon_V (\varphi_i) <<1$.
These possible field values $\varphi_i$ may either all belong to the same interval $\varphi_< < \varphi_i <\varphi_>$, then slow-roll inflation can be realised only in one way (rolling down one hill). Or there are several different, disjoint regions for $\varphi_i$, then the field $\varphi$ may roll down different hills to provide inflation.
For a given $\varphi_i$,  
 all that can be said for generic potentials is that $f_i \equiv f(\varphi_i )$ should be chosen such that $\epsilon (\varphi_i) <<1$ (i.e., $f_i - (1/\sqrt{3}) <<(1/\sqrt{3})$), as well. 
If there exists a choice for $f_i$ giving rise to a realistic inflation scenario at all, however, then the precise choice of the initial value $f_i=f_{i,0}$ is not so important, and neighbouring values $f_i \sim f_{i,0}$, too, will provide inflation. The reason is that different trajectories (different solutions corresponding to different initial values for $f_i$) converge for $\varphi > \varphi_i$, as can be seen easily. Indeed, let us assume that we choose three different solutions $f_{>} $, $f_{0}$, $f_{<}$ defined by  the three initial values $f_{i,>} > f_{i,0} > f_{i,<} >(1/\sqrt{3})$ at $\varphi_i$. It follows immediately from Eq. (\ref{f-root-1}) that, for $\varphi > \varphi_i$, the inequality $f_{>,\varphi} < f_{0,\varphi} < f_{<,\varphi}$ holds, because $\epsilon_V(\varphi_i)$ takes a fixed value at $\varphi_i$, whereas
$\epsilon (\varphi_i)$ is larger for a larger $f(\varphi_i)$. 
On the other hand, the ordering $f_{>} > f_{0} > f_{<}$ is maintained for $\varphi >\varphi_i$ (different trajectories from a one-parameter family of solutions can never intersect or touch). This implies that $f_> $ approaches $f_0$ from above, whereas $f_<$ approaches $f_0$
from below.

We remark that the slow-roll expansion of Section III.B implies a particular choice for the initial value $f_i$ at each order. Indeed, from Eqs. (\ref{W1}) and (\ref{W2}) it easily follows that  
\be
f^{(1)}(\varphi_i) = \frac{1}{\sqrt{3}}, \quad f^{(2)}(\varphi_i) = \frac{1}{\sqrt{3}}\left( 1 + \frac{1}{6}\epsilon_V(\varphi_i) \right) , \quad \ldots
\ee
If the slow-roll expansion converges, then the optimal value for $f_i$ (the one leading to most inflation)  is given by the limiting value
$f_i =\lim_{n\to \infty}f^{(n)}(\varphi_i)$. Good approximations to this optimal trajectory (the "slow-roll attractor") may be achieved already for finite values of $n$.

There exists, however, a specific class of potentials for which the initial condition for the slow-roll attractor can be determined exactly. The first condition which these potentials have to obey is that there exists at least one field value $\varphi_0$ for which $\epsilon_V (\varphi_0) = (U_{,\varphi}/2U)^2(\varphi_0)=0$. The second condition obviously is that the superpotential equation must have a solution in the full interval $\varphi \in [\varphi_0,\varphi_e]$ and, in particular, include the point $\varphi_0$. This implies some further restrictions on the allowed class of potentials \cite{SkT1}, as we shall comment briefly below. Assuming these restrictions, it then follows from the slow-roll expansion of Section III.B that arbitrary orders $\mathcal{W}^{(n)}_{,\varphi}$ or $\epsilon^{(n)}_1$ can be expressed as 
\be
\mathcal{W}^{(n)}_{,\varphi}  = \sqrt{\frac{U}{3}}\frac{U_{,\varphi}}{2U} \left( 1 + \ldots \right) , \quad
\epsilon_{1,{\rm tot}}^{(n)} \equiv  \sum_{k=1}^n \epsilon_1^{(k)} = \epsilon_V \left( 1 + \ldots \right)
\ee  
This implies that, if there exists a field value $\varphi_0$ such that $\epsilon_V (\varphi_0)=0$, then at this field value $\mathcal{W}^{(n)}_{,\varphi}(\varphi_0)=0$ and $\epsilon^{(n)}_1(\varphi_0)=0$ at arbitrary orders. The initial condition for the slow-roll attractor, therefore, is $\epsilon(\varphi_0)=0$, that is, $f(\varphi_0) = (1/\sqrt{3})$, which automatically implies $f_{,\varphi}(\varphi_0)=0$. With the exception of the exponential potential (where the slow-roll attractor can be calculated exactly), all other potentials we consider belong to this restricted class of potentials. We will, therefore, always plot the solution $f$ which corresponds to the slow-roll attractor. For two cases (the exponential and the quadratic potentials) we will, however, also consider additional initial conditions, in order to demonstrate the (fast) convergence to the slow-roll attractor.

Now let us briefly explain the additional restriction on $\mathcal{U}$ implied by the existence of the superpotential  for $\varphi \to \varphi_0$. For simplicity, we assume that $\mathcal{U}(\varphi_0) >0$, such that $\epsilon_V(\varphi_0)=0$ implies $\mathcal{U}_{,\varphi} (\varphi_0)=0$.  The existence of the superpotential at $\varphi_0$ then implies $\mathcal{W}_{,\varphi} (\varphi_0)=0$. Further, the second $\varphi$ derivative of the superpotential equation, evaluated at $\varphi_0$, leads to the algebraic expression
\be
\frac{\mathcal{U}_{,\varphi\varphi}(\varphi_0)}{\mathcal{U}(\varphi_0)} = 2 \frac{\mathcal{W}_{\varphi\varphi}(\varphi_0)}{\mathcal{W}(\varphi_0)} - \frac{2}{3} \left(  \frac{\mathcal{W}_{\varphi\varphi}(\varphi_0)}{\mathcal{W}(\varphi_0)} \right)^2 .
\ee
This expression may be interpreted as an algebraic equation for $\mathcal{W}_{,\varphi\varphi}(\varphi_0)/\mathcal{W}(\varphi_0)$ for a given $\mathcal{U}$. But this equation has a real solution only provided that the "mass" term of the potential $\mathcal{U}$ at $\varphi_0$ obeys the bound
\be \label{SkT-bound}
\frac{\mathcal{U}_{,\varphi\varphi}(\varphi_0)}{\mathcal{U}(\varphi_0)} \le \frac{3}{2},
\ee
as can be checked easily. An equivalent bound, known as the "Breitenlohner-Freedman bound" was originally derived for supergravity in an asymptotic AdS (Anti-deSitter) space \cite{BF1}, as an upper bound on a "tachyon mass". The usual (non-tachyonic) "mass" bound (\ref{SkT-bound}) in standard cosmology was derived, e.g., in \cite{SkT1}.  Similar bounds and restrictions can also be found for the case of  the superpotential (or Hamilton-Jacobi) equation in the multi-scalar-field case, see, e.g., \cite{ACDP} for some recent results. We want to emphasize, however, that these restrictions turn into rigorous bounds only because of the assumption that the superpotential equation can be solved globally, i.e., on full field space. This assumption, in turn, is based on more rigid structures of the underlying field theories, like, e.g.,  supergravity. For the more pragmatic use of the superpotential as a calculational device advocated in the present paper, these restrictions do not necessarily apply, because the existence of the superpotential in a subregion of field space which is sufficient for inflation to occur is all that is required. In particular, the existence of $\mathcal{W}$ at $\varphi_0$ has the additional benefit that it allows to find the slow-roll attractor exactly, but is, in general, not necessary otherwise. Potentials not satisfying the bound (\ref{SkT-bound}) may, therefore, in principle, be viable candidates for inflation that can be treated by the superpotential method, although their exact attractor solutions can be determined only approximately in these cases (because $\mathcal{W}(\varphi)$ cannot be extended to $\varphi_0$).

Still, the fact that the exact slow-roll attractor can be found exactly for the restricted class of potentials discussed above implies that for these potentials the superpotential method has one additional advantage over other methods of slow-roll inflation. In the superpotential formalism, we may simply start the integration of the superpotential equation at $\varphi_0$, thus determining the slow-roll attractor exactly (although, in general, numerically). This is no longer true in the standard FRW approach to scalar field slow-roll inflation. In the FRW evolution, the point $\varphi_0$ must be excluded. The reason is that for $\varphi = \varphi_0$, the inflaton field $\varphi$ (or $\phi$) is not an acceptable time variable, because the transformation from $t$ to $\varphi$, $\dot \varphi = -\kappa^{-1/2}\mathcal{W}_{,\varphi}$, is singular at $\varphi_0$. If one tries, nevertheless, to start the FRW evolution at $\varphi_0$, it may be checked easily that the resulting equations for $\varphi$ are 
$\dot \varphi =0$ and $\ddot \varphi =0$ with the solution $\varphi = \varphi_0 = \mbox{const}$.
In other words, the inflaton field sits at the unstable equilibrium point $\varphi_0$ all the time, and the cosmological evolution is given by the de Sitter expansion. For a nontrivial and physically acceptable solution, one has to start the inflationary evolution at some $\varphi_i$ where $\varphi_0 < \varphi_i <\varphi_e$ (here, it is assumed that inflation ends at $\varphi_e$, i.e., $\epsilon (\varphi_e )=1$).
As a consequence, the exact slow-roll attractor can be determined only approximately.

Let us briefly demonstrate this statement for the generic case. The non-generic case can be treated by a slight modification of our arguments. Here by "generic" we mean that the field value $\phi_0$ where $\epsilon_V(\phi_0)=0$ is finite, $|\phi_0| <\infty$, and that the potential at $\phi_0$ takes a finite, nonzero value, $0<V(\phi_0) <\infty$ (for convenience we momentarily reintroduce the dimensionfull field $\phi$ and potential $V$). This implies that $V_{,\phi}(\phi_0)=0$. If we insert the Hubble function $H$ from Eq. (\ref{Einst1}) into Eq. (\ref{ddot-phi}), we get the following equation in terms of $\phi$ and its derivatives only,
\be \label{eq-phi-only}
\ddot \phi + \sqrt{6\kappa} \sqrt{\left( \frac{1}{2} \dot\phi^2 + V\right)} \, \dot \phi + V_{,\phi} =0.
\ee
This second-order equation has infinitely many solutions, and the exact slow-roll attractor is one of them. But this exact slow-roll attractor cannot be found by starting the integration at $\phi_0$. Indeed, let us assume that we start the integration at some initial time $t_0$ such that $\phi (t_0) = \phi_0$. The condition 
$\epsilon_V(\phi_0) =0$ implies $V_{,\phi}(\phi_0)=0$, by assumption. Further, $\epsilon_V(\phi_0) =0$ implies $\epsilon(\phi_0) =0$ and, therefore, $\dot H (t_0) =0$ and $\dot \phi (t_0) =0$. But the only solution of Eq. (\ref{eq-phi-only}) with these initial conditions is $\ddot \phi =0$, i.e., $\phi = \phi_0 = $ const.
This result is physically obvious if we invoke the analogy of the system (\ref{eq-phi-only}) with the mechanical system of a particle under friction. The initial conditions $V_{,\phi}(\phi_0)=0$ and $\dot\phi (t_0)=0$ imply that the particle in its initial position $ \phi (t_0) =\phi_0$ feels no force and that its initial velocity is zero. But the only possible solution for these initial conditions is $\phi = \phi_0 =$ const. for all times. 
\\
For potentials which obey the constraint (\ref{SkT-bound}), i.e.,
\be \label{SkT-bound-V}
\frac{V_{,\phi\phi}(\phi_0)}{V(\phi_0)} \le \frac{3\kappa}{2}
\ee
for the dimensionful quantities,
even the stronger result holds that the point $\phi_0$ can never be reached in a finite time.
As this is of some independent interest, relating our considerations to the results of \cite{ACDP}, we briefly demonstrate it in appendix B.

We remark that the exact slow-roll attractor (a particular solution of the second-order equation (\ref{eq-phi-only}))
must be distinguished from the leading-order slow-roll attractor. Indeed, the leading slow-roll approximation is equivalent to neglecting the term $(1/2)\dot \phi^2$ in the square root and the term $\ddot \phi$ in equation (\ref{eq-phi-only}), leading to the much simpler first-order equation
\be
\dot \phi = -\frac{V_{,\phi}}{\sqrt{6\kappa V}}.
\ee
Up to time translations, this equation has only one solution, which coincides with the slow-roll attractor in leading order slow-roll. One well-known example is for the quadratic potential $V=M^2\phi^2$ where the leading order attractor is just a constant, 
\be
\dot \phi = -\sqrt{\frac{2}{3\kappa}}M  .
\ee
The exact slow-roll attractor, which can be determined only numerically, approaches this leading order expression for large field values, but describes a curve spiraling towards the point $\phi =0, \dot \phi =0$ in the $(\phi ,\dot \phi)$ plane for small $\phi$, see, e.g., \cite{Muk}, \cite{rem-car}.

\subsection{The exponential potential}
The exponential potential \cite{ab-wi}, \cite{lu-mat}
\be
V(\phi)= \Lambda^4 e^{-\frac{\phi}{m}}
\ee
(here both $\Lambda$ and $m$ are coupling constants with the dimension of mass) has the advantage that it leads to exact solutions for the superpotential and the inflationary observables. It allows, therefore, to study certain features of inflation using simple and exact expressions. On the other hand, it does not lead to a realistic inflationary evolution (e.g., inflation never stops for the exponential potential).  It is still possible that the true inflationary potential may be approximated by the exponential potential during a certain phase of inflation (for a certain interval $\phi \in [\phi_1 ,\phi_2]$ of values of the inflaton field), whereas a different approximation must be used for different (in particular, later) phases. The resulting dimensionless potential (\ref{dim-less}) reads
\be
U(\varphi)= \lambda e^{-\frac{\varphi}{\mu}} \; , \quad \lambda = 2\kappa^2 \Lambda^4 \; , \quad \mu = \sqrt{\kappa}m
\ee
and the dimensionless superpotential equation is
\be
\lambda e^{-\frac{\varphi}{\mu}} = 3\mathcal{W}^2 - \mathcal{W}_{,\varphi}^2.
\ee
Inserting the obvious ansatz $\mathcal{W} = c\exp (-\varphi/2\mu)$ leads to a purely algebraic equation for the real constant $c$ which is solved by
\be \label{W-sol-exp}
c= \sqrt{\frac{\lambda}{3-\frac{1}{4\mu^2}}} \quad \Rightarrow \quad \mathcal{W} = 
\sqrt{\frac{\lambda}{3-\frac{1}{4\mu^2}}} e^{-\frac{\varphi}{2\mu}} .
\ee
The first slow-roll parameter
\be
\epsilon = \frac{\mathcal{W}_{,\varphi}^2}{\mathcal{W}^2} = \frac{1}{4\mu^2} = \frac{U_{,\varphi}^2}{4U^2} = \epsilon_V
\ee
is strictly constant and coincides with the leading slow-roll parameter, which implies $\epsilon_2 =0$, $\delta = -\epsilon$. Inflation therefore never ends, and the exponential potential alone does not provide a realistic scenario. The constant $c$ may be expressed in terms of $\epsilon$ like
\be
c= \sqrt{\frac{\lambda}{3}}\sqrt{\frac{1}{1-\frac{\epsilon}{3}}} = 
\sqrt{\frac{\lambda}{3}}\sqrt{1+\frac{\epsilon}{3} + \left( \frac{\epsilon}{3}\right)^2 + \ldots}
\ee
where we introduced the power series expansion in $\epsilon$, because it is equal to the slow-roll expansion in this simple case. Indeed, from (\ref{W-exp}) we get
\be
\mathcal{W}^{(1)} = \sqrt{\frac{\lambda}{3}} e^{-\frac{\varphi}{2\mu}} \; , \quad 
\mathcal{W}^{(2)} = e^{-\frac{\varphi}{2\mu}} \sqrt{\frac{\lambda}{3}} \sqrt{1 + \beta \frac{\epsilon}{3}}
\ee
and
\be
\mathcal{W}^{(n+1)} = e^{-\frac{\varphi}{2\mu}} \sqrt{\frac{\lambda}{3}} 
\sqrt{1 + \beta \frac{\epsilon}{3}\left( 1 + \beta \frac{\epsilon}{3} \left( 1 + \beta \frac{\epsilon}{3} \ldots \right) \ldots  \right) }
\ee
which, for $\beta =1$, just reproduces the above power series expansion for $c$.

Further, the inflaton solution $\phi(t)$ as well as $H(t)$ and $a(t)$ may be calculated easily. The equation for $\phi$, $\dot \phi = -\kappa^{-\frac{3}{2}}W_{,\phi}$ reads
\be
\dot \phi = -\kappa^{-\frac{3}{2}}\sqrt{\frac{\lambda}{3-\epsilon}} \left( -\frac{1}{2m}\right) e^{-\frac{\phi}{2m}}
\ee
and is solved by 
\be
\phi = m \ln \left( \frac{\epsilon^2\lambda}{\kappa(3-\epsilon)}(t-t_i)^2 \right) 
\ee
where $t_i$ is some arbitrary "initial time" (and, obviously, $t>t_i$). Further,
\be
H(t)= \kappa^{-\frac{1}{2}}W(\phi(t)) = \frac{1}{\epsilon(t-t_i)} 
\ee
and (here $a_*$ is an irrelevant constant)
\be
a(t) = a_* (t-t_i)^\frac{1}{\epsilon}.
\ee
The result for $a(t)$ explains why this model is sometimes called "power-law inflation".

\begin{figure}[h!]
 \centering 
 \includegraphics[width=250pt]{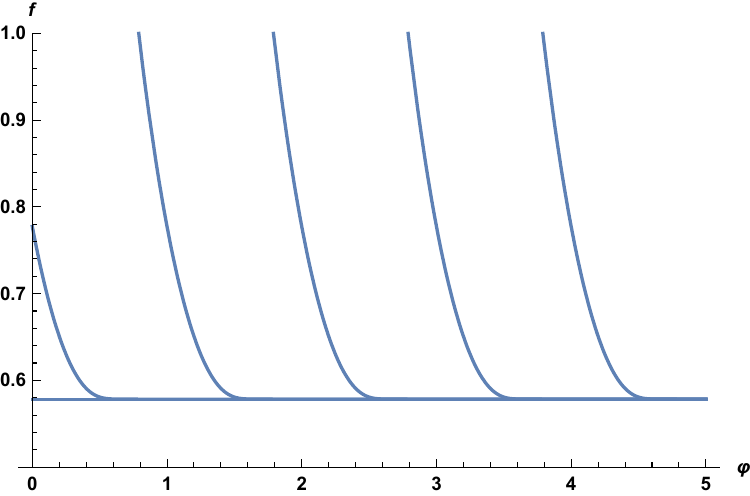} 
\caption{Exponential potential: we plot several numerical solutions to the superpotential equation in terms of the function $f$, in addition to the slow-roll attractor solution $f= (3-\epsilon)^{-1/2}$. Our parameter choice is $\epsilon = 10^{-2}$. The numerical solutions quickly converge to the slow-roll attractor. } 
\label{fig0a}
 \end{figure}

Finally, the exact solution (\ref{W-sol-exp}) obviously is the slow-roll attractor, because $\epsilon_V$ and $\epsilon$ coincide for all $\varphi$. For other solutions, a closed expression cannot be found, although they may be expressed by a one-parameter family of transcendental algebraic equations. We plot some numerical solutions in Fig. \ref{fig0a}, where their fast convergence to the slow-roll attractor can be appreciated.

\subsection{The quadratic potential}
One simple and frequently studied case is the quadratic potential
\be
V= \Lambda^4 \frac{\phi^2}{m^2} \equiv M^2 \phi^2
\ee
which only depends on one parameter (the mass $M=\Lambda^2/m$). The corresponding dimensionless potential is
\be
U(\varphi) = 2M^2\frac{\varphi^2}{\kappa} \equiv \lambda \varphi^2 \; , \quad \lambda = 2M^2 \kappa^{-1}.
\ee
This potential only depends on the overall strength parameter $\lambda$ which does not enter the slow-roll parameters. The model, therefore, provides unique predictions for these parameters.
Before discussing the issues of higher-order slow-roll and full inflationary evolution, we briefly review the inflationary predictions in leading order slow-roll. In leading order slow-roll, the first two potential slow-roll parameters are
\be
\epsilon_V = \frac{1}{4}\frac{U_{,\varphi}^2}{U^2} = \frac{1}{\varphi^2} \; , \quad \delta_V = - \frac{U_{,\varphi\varphi}}{2U} = - \frac{1}{\varphi^2} = - \epsilon_V.
\ee
The (leading slow-roll) condition for inflation to occur, $\epsilon_V<1$, implies $\varphi^2 = \kappa \phi^2 = \phi^2/(2M_{\rm P}^2) >1$, so inflation may happen only for field values above the Planck scale. Inflationary models of this type are sometimes called "large-field models". Using the above results for $\epsilon_V$ and $\delta_V$ (and $\delta \simeq \epsilon_V + \delta_V =0$), together with the observational results for $n_s$ and $r$, we get
\be 
n_s -1 \simeq -4\epsilon_V \simeq - 0.035 \quad \Rightarrow \quad \epsilon_V \simeq 0.00875
\ee
and
\be
r \simeq 16 \epsilon_V \le 0.064 \quad \Rightarrow \quad \epsilon_V \le 0,004 .
\ee
The two results shown above are not compatible, so the purely quadratic potential is strongly disfavored by the recent Planck observations (more conservative estimates for the observables may still slightly reduce this tension). One may, therefore, consider the model just as a simple and instructive case study for the working of inflation.  There is, however, a second and more physical reason for continued interest in the model. It may serve as a "seed" solution for a whole set of "inflationary twin" models of the quadratic potential model, all leading to the same superpotential, inflaton field $\phi(t)$ and inflationary cosmological expansion $a(t)$, while differing in their predictions for some inflationary observables, thus leading to completely viable models of inflation\footnote{Details of these twin models and their use for cosmological inflation will be presented elsewhere.}. 

\begin{figure}[h!]
 \centering 
 \includegraphics[width=250pt]{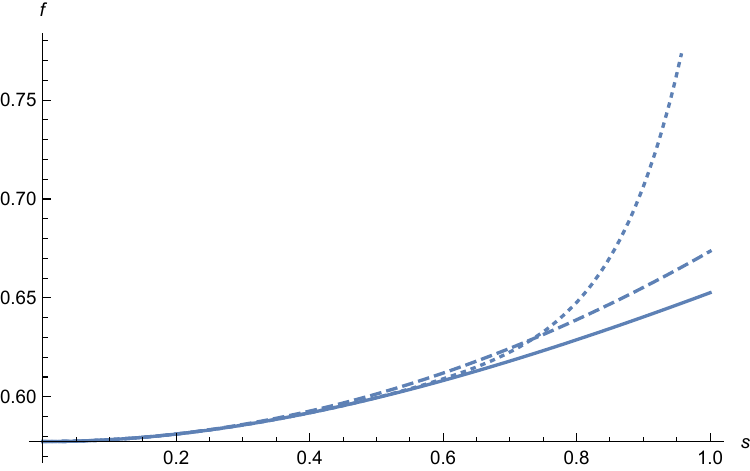} 
\caption{Quadratic potential: we compare the numerical solution for $f(s)$ (continuous line) to the Taylor series expansion up to $s^2$ (dashed line) and up to $s^{10}$ (dotted line). } 
\label{fig1}
 \end{figure}

\begin{figure}[h!]
 \centering 
 \includegraphics[width=250pt]{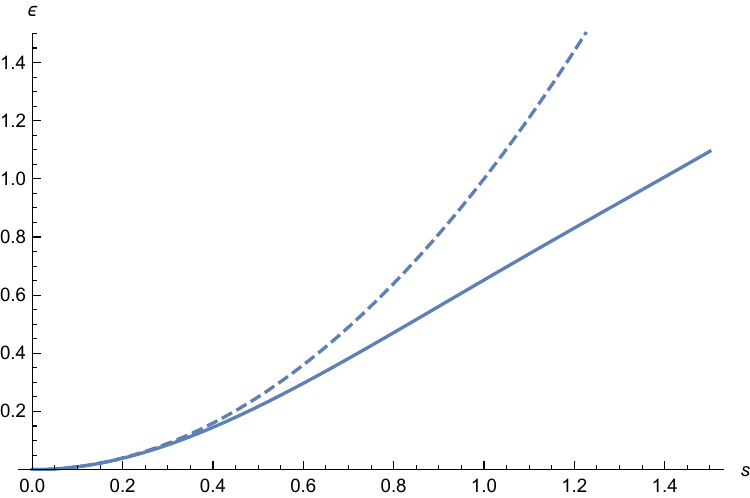} 
\caption{The slow-roll parameter $\epsilon$ for the quadratic potential, in leading order slow-roll (dashed) and from the full numerical integration (continuous line). } 
\label{fig2}
 \end{figure}

If we assume that a given (large) scale $k_*$ leaves the horizon $N_*$ e-folds before the end of inflation, the corresponding evolution in field space for a general model is (here $t_* \equiv t_{k_*}$)
\be
N_* = A(t_e) - A(t_*)  = \int _{\varphi_<}^{\varphi_>} d\varphi \frac{1}{\sqrt{\epsilon}}
\ee
where $\varphi_> = {\rm max}(\varphi_e ,\varphi_*)$,  $\varphi_< = {\rm min}(\varphi_e ,\varphi_*)$ and $\varphi_* = \varphi (t_*)$. In leading order slow-roll, $\epsilon$ is replaced by $\epsilon_V$, and $\varphi_e$ is defined by $\epsilon_V (\varphi_e)=1$. Assuming now $N_* = 60$, we get for the quadratic potential (where $\epsilon_V = 1/\varphi^2$ and $\varphi_e =1$),
\be
60 = \int_1^{\varphi_*} d\varphi \varphi = \frac{1}{2}\left( \varphi_*^2 -1 \right) \quad \Rightarrow
\quad \varphi_* = 11 \; , \quad \phi_* = 11 \sqrt{2} M_{\rm P} \simeq 15 M_{\rm P}.
\ee
We will see in a moment how these results change beyond slow-roll. 
For this, we use the equation for $f$ introduced in (\ref{f-eq}).
In particular, for the quadratic potential $U=\lambda \varphi^2$ we get
\be
1 = 3f^2 - \beta \left(\frac{1}{\varphi}f + f_{,\varphi} \right)^2
\ee
or, introducing the new variable $s=\varphi^{-1}$,
\be \label{f-s-eq}
1 = 3f^2 - \beta s^2 \left( f - sf_{,s} \right)^2  .
\ee
In principle, we should choose the root (\ref{f-eq-root}) with the plus sign, but it turns out that, in this case, Eq. (\ref{f-s-eq}) with the initial condition $f(s=0)=(1/\sqrt{3})$ only has one solution, corresponding to the correct sign.
Further, from expression (\ref{f-s-eq}) it is obvious that, for the quadratic potential, the slow-roll expansion (the formal power series expansion in $\beta$) is completely equivalent to a Taylor series expansion of $f$ in powers of $s^2$.  An expansion up to fifth order in $s^2$, e.g., leads to
\be
f(s) = \frac{1}{\sqrt{3}}\left( 1 + \frac{1}{6}\beta s^2 - \frac{5}{72}\beta^2 s^4 + \frac{37}{432}\beta^3 s^6- \frac{197 }{1152}\beta^4 s^8+ \frac{28895}{62208}\beta^5 s^{10}-+ \ldots \right) .
\ee
A tendency which becomes much more pronounced for even higher orders is that the numerator of the expansion terms grows faster than the denominator (already for $\beta^6$ the numerator is bigger than the denominator). This implies that, for $\beta =1$, the Taylor series expansion will not converge up to $s=1$ but, rather, have a smaller radius of convergence. In Fig. \ref{fig1}, we compare the power series expansion to the numerical solution of Eq. (\ref{f-s-eq}), both of which can be calculated with simple Mathematica routines. 
For the numerical integration, we start at the very small value $s_0 = 10^{-4}$ with the quadratic approximation $f(s_0)= (1/\sqrt{3})(1 + s_0^2/6)$.  The expansion up to $s^{10} $ starts to deviate from the numerical solution at about $s=0.5$, and even higher orders confirm that the radius of convergence is close to that value (higher orders do not improve the convergence for $s>0.5$, because the Taylor series is alternating).
Fig. \ref{fig1} also seems to indicate that the next-to-leading slow-roll approximation works quite well in the whole interval $s \in [0,1]$. In a moment we will see, however, that higher than leading-order slow-roll results improve the precision for small $s$ but are not useful outside the radius of convergence. For $s\sim 1$, numerical integration is required. As an example, we plot in Fig. \ref{fig2} the first slow-roll parameter $\epsilon$, both in leading order slow-roll ($\epsilon_V$) and using the numerical solution for $f(s)$, where $\epsilon$ is defined in terms of $f$ like
\be
\epsilon = s^2 \left( 1-s\frac{f_{,s}}{f} \right)^2 .
\ee
Higher-order slow-roll results improve the convergence for small $s$ but do not make too much sense for larger $s$. The next-to-leading order result
\be
\epsilon_1^{(2)} = s^2\left( 1-\frac{2}{3}s^2 \right) ,
\ee
e.g., never reaches the value $\epsilon =1$, so inflation does not seem to end in this approximation.
Near the end of inflation, Fig. \ref{fig2} shows a rather big difference between leading-order slow-roll and the full numerical solution. The field value where inflation ends (where $\epsilon =1$), e.g., is $s_e =1$ in slow-roll but $s_e \simeq 1.393$ in the full numerical solution. 

The field value $s_*$ (i.e., $\varphi_*$) at which a sufficiently large scale $1/k_*$ (with e-fold number $N_*$) leaves the horizon, however, only very weakly depends on these differences. The reason is that $N_* = \int d\varphi \epsilon^{-1/2}$ and, therefore, receives almost all its contributions from regions of small $\epsilon$, whereas differences in the region $\epsilon \sim 1$ are not very relevant. Concretely,
\be
N_* = \int _{s_*}^{s_e} \frac{ds}{s^3}\frac{f}{f-sf_{,s}}
\ee
and for $N_* = 60$, $s_*$ in leading order (where $s_e =1$ and $f=$ const) is $s_* = (1/11) = 0.0909$ corresponding to $\varphi_* = 11$, whereas for the full numerics (where $s_e = 1.393$), $s_* = 0.0917$ corresponding to $\varphi_* = 10.905$. This implies that also the values of $\epsilon$ at this scale are quite similar, namely  $\epsilon_{V}(\varphi_*=11) = (1/121) =0.008264$ and $\epsilon (\varphi_*=10.905)=  0.008362$. 

\begin{figure}[h!]
 \centering 
 \includegraphics[width=250pt]{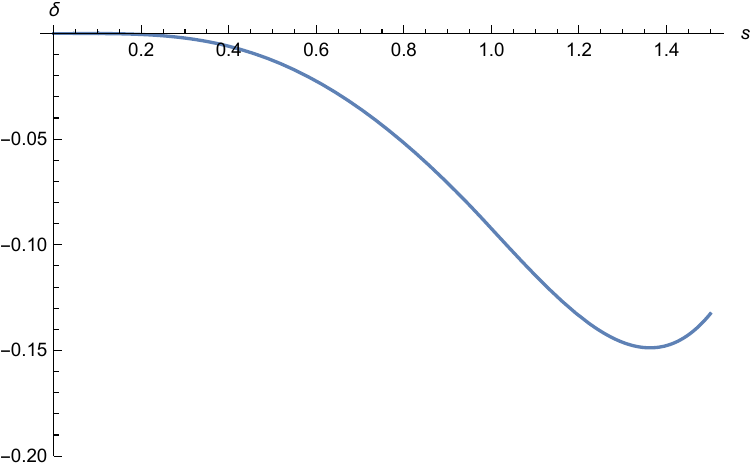} 
\caption{The slow-roll parameter $\delta$ for the quadratic potential from the full numerical integration. } 
\label{fig3}
 \end{figure}

Finally, for the second slow-roll parameter $\delta$ for a general potential, we calculate
\be
\delta = -\frac{\mathcal{W}_{\varphi\varphi}}{\mathcal{W}} = \delta_V + \epsilon_V - \frac{U_{,\varphi}}{U}\frac{f_{,\varphi}}{f} - \frac{f_{,\varphi\varphi}}{f}.
\ee
For the quadratic potential, the first two terms at the r.h.s. cancel. Using, further, the new variable $s=\varphi^{-1}$, we get the simple result
\be
\delta = -s^4\frac{f_{,ss}}{f}
\ee
which is strongly suppressed for small $s$, as it must be. We show the numerical result for $\delta$ in Fig. \ref{fig3}.
In particular, at $\varphi_* = 10.905$, $\delta$ takes the value $\delta(\varphi_*)= -0.0000231$, that is, it is very small, as expected.

\begin{figure}[h!]
 \centering 
 \includegraphics[width=250pt]{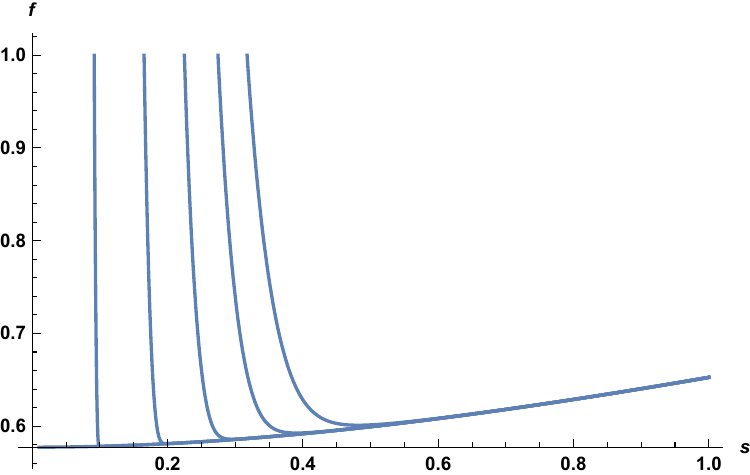} 
\caption{Quadratic potential: we show several numerical solutions for $f(s)$, in addition to the slow-roll attractor solution obeying the initial condition $f(s=0) = 1/\sqrt{3}$. As can be seen, these solutions quickly converge to the slow-roll attractor. In addition, it follows from Eq. (\ref{f-s-eq}) that a solution which does not obey the initial condition $f(s=0) = 1/\sqrt{3}$ must diverge in the limit $s\to 0$. Also this divergent behaviour can clearly be seen in the figure. } 
\label{fig3a}
 \end{figure}
As explained in Section V.B, the solution $f(s)$ of Eq. (\ref{f-s-eq}) shown in Fig. \ref{fig1} is the slow-roll attractor. In Fig. \ref{fig3a} we show several numerical solutions which do not obey the slow-roll attractor initial condition  $f(s=0) = 1/\sqrt{3}$, in order to demonstrate their fast convergence to the slow-roll attractor.

\subsection{The hill-top potential}
In hill-top models of inflation it is assumed that the inflaton field starts its roll-down very close to a local maximum (the "top of a hill"). One classic example would be the "Mexican hat" potential of spontaneous symmetry breaking, but arguably the simplest of all possible cases (and the one we shall consider here) is given by the potential
\be
V = \Lambda^4 \left( 1-\frac{\phi^2}{m^2}\right) .
\ee
This potential as it stands cannot be correct for all values of $\phi$, because it becomes negative for $|\phi | >m$. The idea is that the above expression provides the leading terms in a power series expansion about the maximum at $\phi =0$, and that, for the purpose of inflation, it is sufficient to consider these leading terms. In any case, inflation can never occur for negative potentials (a negative "effective vacuum energy") so inflation is guaranteed to end for $|\phi | <m$. The corresponding dimensionless potential is
\be
U = \lambda \left( 1 - \frac{\varphi^2}{\mu^2} \right) \; , \quad \lambda = 2\kappa^2\Lambda^4 \; , \quad \mu = \sqrt{\kappa} m.
\ee
Up to the overall strength parameter $\lambda$, the model depends on one parameter $\mu$. It will, therefore, provide a one-parameter family of predictions for the slow-roll parameters.

Let us, again, start by reviewing the leading slow-roll behaviour. The two potential slow-roll parameters are
\be
\epsilon_V = \frac{1}{\mu^2} \frac{\left( \varphi/\mu \right)^2}{\left(1 - \left(\varphi /\mu \right)^2\right)^2}
\; , \quad \delta_V =  \frac{1}{\mu^2} \frac{1}{1 - \left(\varphi /\mu \right)^2}
\ee
so while $\epsilon_V$ may be small for any value of $\mu$ (by choosing a sufficiently small $\varphi$), the smallness of $\delta_V$ requires $\mu $ to be large. The condition $\epsilon_V (\varphi_e)=1$ for inflation to end is solved by
\be
\varphi_e = \sqrt{\mu^2 + \frac{1}{4}} - \frac{1}{2}
\ee
and the expression for the number of e-folds is
\be
N_* = \int_{\varphi_*}^{\varphi_e} d\varphi \epsilon_V{}^{-1/2} = \left( \mu^2 \ln \varphi - \frac{\varphi^2}{2} \right)_{\varphi_*}^{\varphi_e}
\ee
and may be solved numerically for particular choices for $\mu$ and $N_*$. E.g., for $\mu =9$, $N_* = 60$ we find $\varphi_e =8.514 $ and $\varphi_* =2.718 $, that is, again field values above the Planck scale. Small $\mu$ would lead to small field values, but as said, small $\mu$ lead to unacceptably large values for $\delta_V$. Large (but not too large) values of $\mu$, on the other hand, can provide results in complete agreement with the observational bounds. The case $N_* =60$, $\mu =9$, e.g.,  leads to $\epsilon_V(\varphi_*)=0.00136$ and $\delta_V(\varphi_*)=0.01494$ and, therefore, to $n_s -1 = - 0.03532$ and $r = 0.02176$.

Beyond slow-roll, Eq. (\ref{f-eq}) for the hill-top potential reads
\be
1 = 3f^2 -\beta \left( -\frac{1}{\mu^2} \frac{\varphi}{1-(\varphi^2/\mu^2)}f + f_{,\varphi} \right)^2.
\ee
Firstly, it is useful to introduce the new variable $s=\varphi/\mu$ (that is, measure the field $\phi$ in units of $m$ rather than $\kappa^{-1/2}$), because then $\mu^{-2}$ becomes a common prefactor of the parenthesis. So, the physical parameter $\mu^{-2}$ plays the role of the formal expansion parameter $\beta$, and we may set $\beta =1$ without losing any information, resulting in
\be
1 = 3f^2 - \frac{1}{\mu^2} \left( -\frac{s}{1-s^2}f + f_{,s} \right)^2.
\ee
As $\mu^{-2}$ now plays the role of the slow-roll expansion parameter, this implies that the slow-roll expansion will work particularly well in the region of large $\mu$, which is precisely the region of physical interest. The numerical calculation will completely confirm this expectation.
Further, a Taylor expansion $f=\sum f_{2n}s^{2n}$ may again be performed, but in this case the Taylor expansion is {\em not} equivalent to a formal power series expansion in $\beta$ (or $\mu^{-2}$). Still, we expect good convergence of this Taylor expansion for large $\mu$.
Secondly, we assume $0< \varphi_* < \varphi_e$ for the field range of inflation, such that $\varphi$ grows with inflation, implying that the minus sign must be chosen in (\ref{f-eq-root}),
\be
f_{,s} = \frac{s}{1-s^2}f - \mu \sqrt{3f^2 -1}.
\ee
Our initial condition is $f(s=0)=1/\sqrt{3}$. 
We shall perform the explicit numerical integration for $\mu = 9$, because this value leads to inflationary predictions in complete agreement with the observational constraints. 
\begin{figure}[h!]
 \centering 
 \includegraphics[width=250pt]{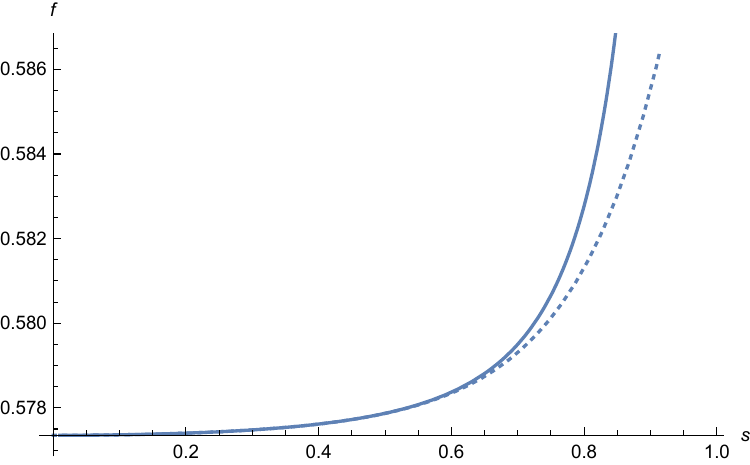} 
\caption{Hill-top potential: the numerical solution for $f(s)$ (continuous line) and the Taylor expansion up to fifth order in $s^2$ (dotted line). } 
\label{fig4}
 \end{figure}
In Fig \ref{fig4} we show the result of the numerical integration together with a Taylor expansion up to order $s^{10}$. We do not display the Taylor coefficients $f_{2n}$ in this case, because the resulting expressions are rather lengthy. Firstly, we see that the Taylor expansion approximates the true solution very well up to $s\sim 0.6$. However, in this case all Taylor coefficients $f_{2n}$ are positive, so going to even higher orders will make the Taylor expansion grow faster for larger $s$ and approximate the true solution even better. That is to say, the radius of convergence will be close to $s=1$. Secondly, $|f-1/\sqrt{3}|$ is very small in the whole displayed interval, which means that the slow-roll expansion is trustworthy and that already the leading slow-roll results should be rather good.

\begin{figure}[h!]
 \centering 
 \includegraphics[width=250pt]{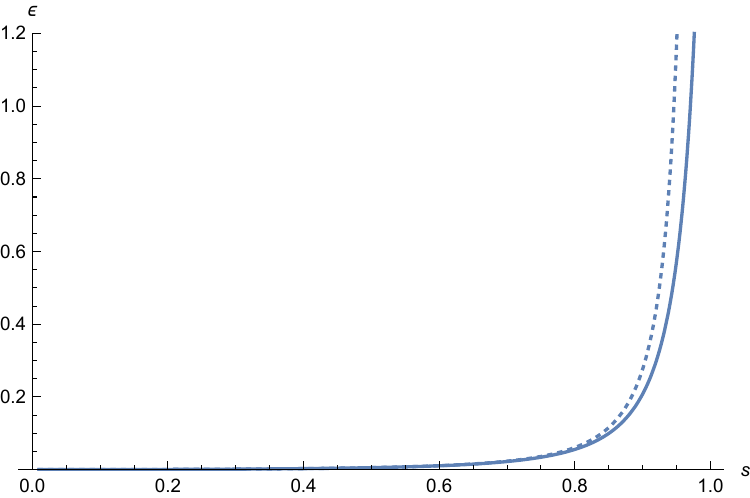} 
\caption{The slow-roll parameter $ \epsilon$ for the hill-top potential, for the full numerical solution for $f(s)$ (continuous line) and for $f=1/\sqrt{3}$ (leading slow-roll $\epsilon_V$, dotted line). } 
\label{fig5}
 \end{figure}

\begin{figure}[h!]
 \centering 
 \includegraphics[width=250pt]{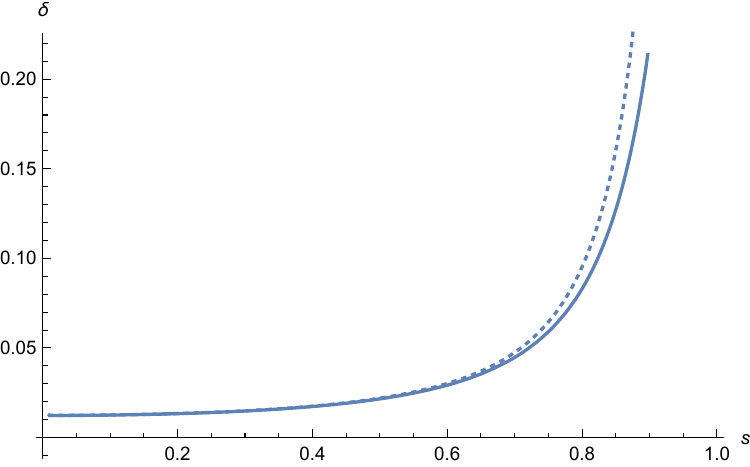} 
\caption{The slow-roll parameter $ \delta$ for the hill-top potential, for the full numerical solution for $f(s)$ (continuous line) and for $f=1/\sqrt{3}$ (leading slow-roll, $\delta |_{\rm lsr} = \delta_V + \epsilon_V$, dotted line). } 
\label{fig6}
 \end{figure}

This is precisely what happens. The slow roll parameter $\epsilon$, for instance, reads
\be
\epsilon = \frac{1}{\mu^2} \left( -\frac{s}{1-s^2} + \frac{f_{,s}}{f} \right)^2
\ee
and $\epsilon_V$ is given by the same expression, but for $f_{,s}=0$. For $\mu =9$, $\epsilon$ and $\epsilon_V$ are shown in Fig. \ref{fig5}.
They are almost indistinguishable up to $s\sim 0.8$, and quite similar in the whole displayed interval. E.g., the end of inflation $\epsilon (s_e)=1$ in the exact numerical case happens for $s_e =0.9703$, whereas in leading slow-roll $\epsilon_V(s_e)=1$ occurs at $s_e = 0.9460$. 
The e-fold number is
\be
N_* = \int_{\varphi_*}^{\varphi_e} d\varphi \epsilon^{-1/2} = \mu^2 \int_{s_*}^{s_e} ds \left( -\frac{s}{1-s^2} + \frac{f_{,s}}{f} \right)^{-1}
\ee
(in leading slow-roll the same expression with $f_{,s}=0$ is valid). For $N_* =60$, the field value $s_*$ where the corresponding scale crosses the horizon is $s_* = 0.3068$ for the exact numerical calculation and $s_* = 0.3017$ in leading slow-roll. The second slow-roll parameter $\delta$,
\be
\delta(s) = \frac{1}{\mu^2} \left( \frac{1}{1 - s^2} + \frac{s^2}{(1 - s^2)^2} + 
    2 \frac{s}{1 - s^2} \frac{f_{,s} }{f } - 
    \frac{f_{,ss} }{f }\right) ,
\ee
 is shown in Fig. \ref{fig6}, both for the exact numerical and the leading slow-roll calculation. 
Finally, the values of $\epsilon$ and $\delta$ at $s_*$ are $\epsilon(s_*)=0.00140$ and $\delta (s_*)=0.01493$ for the exact numerical calculation and $\epsilon(s_*)=0.00136$ and $\delta(s_*)=0.01494$ in leading order slow-roll. The exact and leading slow-roll results are very close to each other and in complete agreement with all observational bounds.

If we assume, instead, that large scales leave the horizon 50 e-folds before the end of inflation, i.e., $N_* =50$, then it turns out that the value of the parameter $\mu$ which best agrees with the constraints from observations is $\mu = 11$. We gather the results in this case, together with the case $N_* = 60$, $\mu =9$ already considered, in Table \ref{tab1}.

\begin{table}[htb]
\caption{The slow-roll parameters $\epsilon_*\equiv \epsilon (s_*)$ and $\delta_*\equiv \delta (s_*)$ evaluated at $s_*$ for the hill-top potential, for the parameter values $\mu = 9$, $N_* = 60$ and $\mu = 11$, $N_*=50$ (lsr = leading slow-roll).}
\begin{tabular}{c|cccccccc}
\hline\hline
\qquad & \quad  $s_e$ \qquad & \quad $s_e$  \qquad & \quad $s_*$   \qquad & \quad $s_*$ \qquad & \quad $\epsilon_*$ \qquad  & \quad $\epsilon_*$ \qquad &\quad $\delta_*$ \qquad &\quad $ \delta_*$ \\
\qquad & \quad  exact \qquad & \quad lsr \qquad & \quad exact   \qquad & \quad lsr \qquad & \quad exact \qquad  & \quad lsr \qquad &\quad exact \qquad &\quad lsr \\

\hline
    $\mu = 9$, $N_* = 60$    &         0.9703   &    0.9460     &    0.3068    &    0.3017   &  0.00140 &  0.00136     &  0.01493 & 0.01494 \\
       $\mu = 11$,  $N_*=50$  &     0.9755    &    0.9556  &       0.4465  &       0.4414    & 0.00254 &   0.00248    & 0.01277 & 0.01275 \\
       
\hline      
\hline       
\end{tabular}
\label{tab1}
\end{table}

\subsection{The Starobinsky potential}
The Starobinsky model of inflation \cite{starob1} is based on the purely gravitational action
\be \label{star-act}
S_{\rm St} = \frac{1}{\kappa} \int d^4 x \sqrt{|g|}\left( R + \frac{1}{6M_{\rm St}^2}R^2 \right)
\ee
where the $R^2$ term is interpreted as a leading order correction due to quantum gravity effects.  In the absence of a generally accepted theory of quantum gravity, $M_{\rm St}$ is treated as a free parameter. This action may be brought into the form of the standard Einstein-Hilbert (EH) action plus a scalar field by the following steps. Firstly, an auxiliary field $\chi$ is introduced to provide the action
\be 
S = \frac{1}{\kappa} \int d^4 x \sqrt{|g|} \left( \left( 1+\frac{1}{3M_{\rm St}^2}\chi \right) R - \frac{1}{6M_{\rm St}^2}\chi^2 \right) .
\ee
The purely algebraic field equation $\chi = R$ satisfied by $\chi$ reproduces the original Starobinsky action. This action is now linear in the curvature scalar $R$, but the term linear in $R$ is not normalised like the EH term (the action is expressed in a "Jordan frame"). The correct normalisation may be achieved by a ($\chi$-dependent) Weyl transformation of the metric to an Einstein frame. The same Weyl transformation acting on $R$ induces a kinetic term for the scalar field $\chi$ but, again, not with the standard normalisation for this term. The canonical normalisation for the kinetic term of $\chi$ may be achieved by a field transformation $\chi \to \phi$. The result is the standard EH plus scalar field action (\ref{action}) with the Starobinsky potential
\be
V = \Lambda^4 \left( 1 - e^{-\frac{\phi}{m}}\right)^2
\ee
where $\Lambda$ is related to $M_{\rm St}$. Further, when this potential is derived from the Starobinksy action (\ref{star-act}) as indicated, then the parameter $m$ takes the fixed (Starobinsky) value $m = m_{\rm St} = \sqrt{3/2} M_{\rm P} = \sqrt{3/(4\kappa)}$. We shall treat $m$ as a free parameter in the general discussion but choose the Starobinsky value for the explicit numerical evaluation. The corresponding dimensionless potential is
\be
U = \lambda \left(1-e^{-\frac{\varphi}{\mu}}\right)^2 \; , \quad \lambda = 2\kappa^2\Lambda^4 \; , \quad \mu = \sqrt{\kappa} m
\ee
and the Starobinsky value for $\mu$ is $\mu_{\rm St}= \sqrt{3/4}$.
The potential slow-roll parameters are
\be
\epsilon_V = \frac{e^{-\frac{2\varphi}{\mu}}}{\mu^2 \left( 1-e^{-\frac{\varphi}{\mu}}\right)^2}
\; , \quad \delta_V = \frac{e^{-\frac{\varphi}{\mu}}\left(1-2 e^{-\frac{\varphi}{\mu}}\right)}{\mu^2 \left( 1-e^{-\frac{\varphi}{\mu}}\right)^2}
\ee
$\epsilon_V$ becomes zero in the limit $\varphi\to+\infty$, so $\varphi$ diminishes during inflation and the plus sign must be chosen in Eq. (\ref{f-eq-root}). Further, it is useful to introduce the new variable $s = \exp (-\varphi /\mu )$. The potential slow-roll parameters may then be expressed like
\be
\epsilon_V = \frac{s^2}{\mu^2 (1-s)^2} \; , \quad \delta_V =\frac{s-2s^2}{\mu^2 (1-s)^2},
\ee
inflation in leading slow-roll ends at
\be
\epsilon_V =1 \quad \Rightarrow \quad s_e = \frac{\mu}{1+\mu}
\ee
and the number of e-folds is
\be
N_* = -\int_{\varphi_*}^{\varphi_e} d\varphi \epsilon_V^{-\frac{1}{2}} = \mu \int_{s_*}^{s_e}\frac{ds}{s}\epsilon_V^{-\frac{1}{2}} = \mu^2 \left( -\frac{1}{s} - \ln s \right)_{s_*}^{s_e}.
\ee
For the Starobinsky value, $\mu = \mu_{\rm St}$ is of order one. A large $N_* \sim 50-60$, therefore, requires a small $s_*$ and permits the (somewhat rough but frequently used) approximation $N_* \simeq \mu_{\rm St}^2/s_*$, and the corresponding approximations $\epsilon_V(s_*) \simeq \mu_{\rm St}^2/N_*^2 = (3/4)N_*^{-2}$
and $\delta_V(s_*) \simeq N_*^{-1}$.

Going beyond leading slow-roll, the superpotential equation (\ref{f-eq}) for $f(s)$ reads (we set $\beta =1$)
\be
 1 = 3f^2 -\frac{s^2}{\mu^2} \left( -\frac{1}{1-s} f + f_{,s} \right)^2 
 \ee 
 and $f$ obeys the initial condition $f(s=0)= 1/\sqrt{3}$. As in the case of hill-top inflation, the physical parameter $\mu^{-2}$ plays, at the same time, the role of the formal expansion parameter $\beta$. In the present case, however, the physically relevant value $\mu_{\rm St}$ is not large, therefore we expect significant differences between leading slow-roll and exact results, except for very small values of $s$. The numerical calculations fully confirm this expectation. 
\begin{figure}[h!]
 \centering 
 \includegraphics[width=250pt]{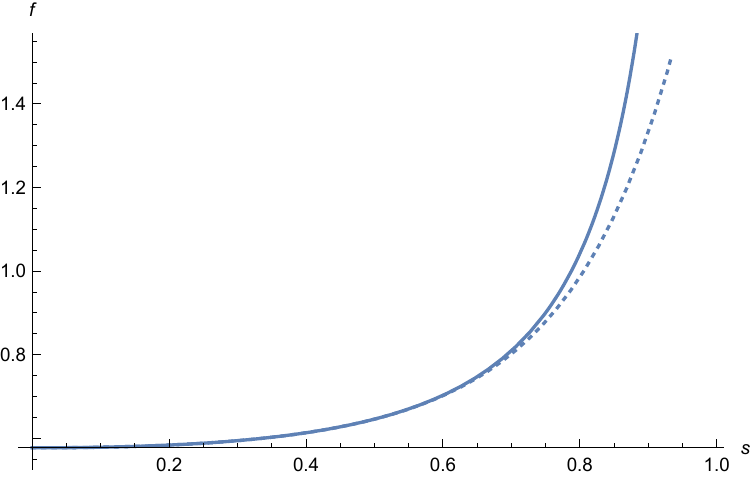} 
\caption{Starobinsky potential: the numerical solution for $f(s)$ (continuous line) and the Taylor expansion up to order $s^{10}$ (dotted line). } 
\label{fig7}
 \end{figure}
We plot the numerical result for $f$ as well as a Taylor expansion $f=\sum_n f_n s^n$ up to order $s^{10}$ (in this case both even and odd orders contribute) in Fig. \ref{fig7}. Good convergence of the Taylor series expansion can be seen. On the other hand, $f$ deviates from its minimum value $1/\sqrt{3}$ appreciably rather soon, therefore we expect noticeable differences between the leading slow-roll and the exact results for the slow-roll parameters. 
The explicit expressions for the slow-roll parameters are
\be
\epsilon = \frac{s^2}{\mu^2}\left( -\frac{1}{1-s} + \frac{f_{,s}}{f}\right)^2
\ee
and
\be
\delta = \frac{1}{\mu^2} \left( \frac{s}{1-s} + \frac{3s^2 -s}{1-s}\frac{f_{,s}}{f} -s^2 \frac{f_{,ss}}{f} \right)
\ee
\begin{figure}[h!]
 \centering 
 \includegraphics[width=250pt]{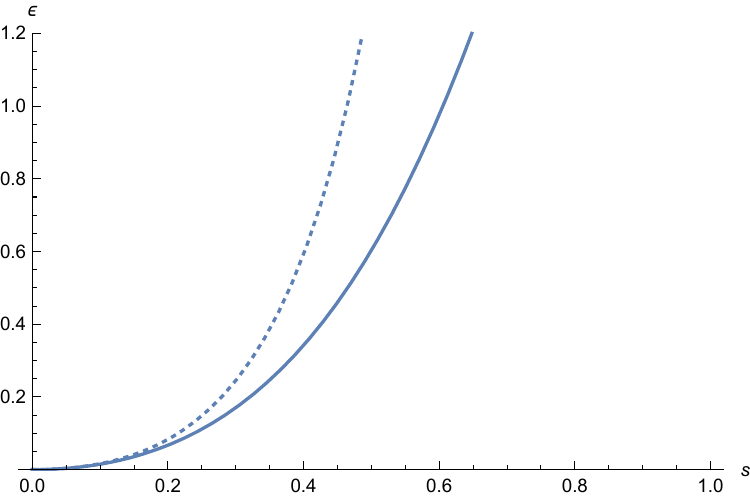} 
\caption{The slow-roll parameter $ \epsilon$ for the Starobinsky potential, for the full numerical solution for $f(s)$ (continuous line) and for $f=1/\sqrt{3}$ (leading slow-roll $\epsilon_V$, dotted line). } 
\label{fig8}
 \end{figure}
\begin{figure}[h!]
 \centering 
 \includegraphics[width=250pt]{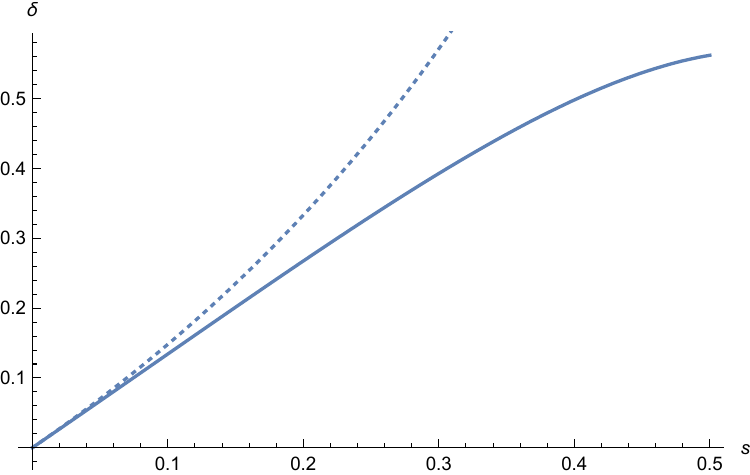} 
\caption{The slow-roll parameter $ \delta$ for the Starobinsky potential, for the full numerical solution for $f(s)$ (continuous line) and for $f=1/\sqrt{3}$ (leading slow-roll, $\delta |_{\rm lsr} = \delta_V + \epsilon_V$, dotted line). } 
\label{fig9}
 \end{figure}
and we plot both the leading slow-roll results and the full numerical results in Figs. \ref{fig8} and \ref{fig9}.
The differences between leading slow-roll and the exact numerical result are clearly visible both for $\epsilon$ and for $\delta$. 

\begin{table}[htb]
\caption{The slow-roll parameters $\epsilon_*\equiv \epsilon (s_*)$ and $\delta_*\equiv \delta (s_*)$ evaluated at $s_*$ for the Starobinsky potential, for the Starobinsky value $\mu = \mu_{\rm St} = \sqrt{3/4}$, and for the e-fold numbers $N_* = 60$, $N_*=56$ and $N_*=50$ (lsr = leading slow-roll).}
\begin{tabular}{c|cccccccc}
\hline\hline
\qquad & \quad  $s_e$ \qquad & \quad $s_e$  \qquad & \quad $s_*$   \qquad & \quad $s_*$ \qquad & \quad $\epsilon_*$ \qquad  & \quad $\epsilon_*$ \qquad &\quad $\delta_*$ \qquad &\quad $ \delta_*$ \\
\qquad & \quad  exact \qquad & \quad lsr \qquad & \quad exact   \qquad & \quad lsr \qquad & \quad exact \qquad  & \quad lsr \qquad &\quad exact \qquad &\quad lsr \\

\hline
    $N_* = 60$    &         0.6054   &    0.4641     &    0.01193    &    0.01165   &  0.000192 &  0.000185     &  0.01593 &  0.01572 \\
    $N_* = 56$    &         0.6054   &    0.4641     &    0.01275    &    0.01243   &  0.000220 &  0.000211    &  0.01703 &  0.01678 \\
         $N_*=50$  &     0.6054    &    0.4641  &       0.01421  &       0.01382    & 0.000274 &   0.000262    & 0.01898 & 0.01869 \\
       
\hline      
\hline       
\end{tabular}
\label{tab2}
\end{table}
Finally, in Table \ref{tab2} we display the slow-roll parameters for several values of $N_*$. For typical e-fold numbers $50 \le N_* \le 60$, differences between exact and leading slow-roll Hubble flow parameters are of the order of a few percent. Further we find that, for $N_*=56$, the calculated values for $\epsilon_*$ and $\delta_*$ precisely match the observed value of the scalar spectral index, $n_s - 1 = -2\delta_* -4\epsilon_* = -0.03494$.  Also the bound on the tensor-to-scalar ratio $r=16\epsilon_*$ is easily matched, because $\epsilon_*$ results very small in Starobinsky inflation. This demonstrates that Starobinsky inflation is still a viable model of cosmological inflation.

\section{Summary}
It was the main purpose of this paper to demonstrate the simplicity and usefulness of the superpotential method in cosmological inflation. In the particular case of single-field inflation, the superpotential equation is a simple first-order ODE which is completely equivalent to the standard evolution equations of single-field inflation. 
First of all, the superpotential method allows to calculate the higher-order Hubble flow functions in the slow-roll expansion in terms of a simple formal power series expansion of the superpotential equation.
Secondly, for a large class of physically well-motivated potentials, it allows to exactly determine the slow-roll attractor to which generic inflationary solutions converge.
It also implies a rigorous upper bound on the first Hubble flow function, $0\le \epsilon \le 3$.
Further, all inflationary observables may be calculated exactly (beyond the slow-roll approximation) from the solutions of the superpotential equation. These solutions, in turn, may be found, e.g., with the help of a simple Mathematica program.  

In addition, both the superpotential equation and its solutions allow for an easy and direct estimate of the reliability of the slow-roll expansion. On the one hand, for a large class of inflaton potentials $V$ it holds that they depend on the inflaton field $\phi$ only via the combination $\phi/m = \varphi/\mu$, where $m$  is a mass scale, and $\varphi = \sqrt{\kappa} \phi$, 
$\mu = \sqrt{\kappa} m$ are their dimension-less versions. In all these cases, the physical parameter $\mu^{-2}$ plays the role of the formal slow-roll expansion parameter in the superpotential equation, such that the leading slow-roll approximation will already provide very good results for large $\mu$ (that is, small $\mu^{-2}$).
If, instead, the physically relevant value of $\mu$ is not large (small or of order one), like e.g. in the case of Starobinsky inflation, then noticeable differences between leading slow-roll and the full solution have to be expected. On the other hand, if a superpotential solution $\mathcal{W}$ of the superpotential equation [and the corresponding auxiliary function $f = (1/\sqrt{U}) \mathcal{W}$] is known, then slow-roll is reliable as long as $f$ obeys the inequality $|f-1/\sqrt{3}|<<1/\sqrt{3}$. Further, for a wide class of potentials a Taylor series expansion of the auxiliary function $f$ is available in addition to the numerical solution. The radius of convergence of this Taylor series is, again, related to the accuracy of the slow-roll expansion.

We treated some particular examples of single-field inflationary models to explicitly demonstrate the simplicity of the method in calculating inflationary observables. We think that, owing to its simplicity, one of the main virtues of the method is in the realm of model building for inflationary dynamics. If a particular model of inflation is encountered as the result of some theoretical investigation, then some elementary knowledge of Mathematica and at most a few hours of dedication are all that is needed to calculate all its inflationary predictions even exactly (i.e., beyond slow-roll), along the lines of the specific examples considered here.  
As a matter of fact, the field of cosmological inflation and inflationary calculations is well-developed, and there are even on-line resources available to assist these calculations. We believe, nevertheless, that the methods presented here may prove very useful also from a more practical point of view. First of all, if all that one wants to find are the inflationary predictions of a specific model,  
it would imply significantly more effort to turn on all that machinery, instead of just doing the simple calculations required by the superpotential formalism. Secondly, the numerical evaluation based on the superpotential is faster (requires less computing time) than the one based on the FRW evolution, because the superpotential equation is just one first-order ODE.  This gain in computing time is not relevant if just one specific model is considered, but it may become important for the statistical inference of the observationally most viable models from a large set of initial models and parameter values, as was done, e.g., in \cite{enzy}, \cite{best}.

To sum up, the superpotential method allows an easy estimate of the precision of the slow-roll approximation, on the one hand. It also provides a simple and systematic procedure to calculate higher orders of the slow-roll expansion. On the other hand, it is a simple tool for inflationary calculations beyond the slow-roll approximation. In our concrete examples, we considered inflationary models which give rise to slow-roll inflation even beyond any slow-roll approximation (i.e., using the exact inflationary trajectories instead of the slow-roll expansion), that is to say, which lead to small values for the Hubble flow functions in the interval in field space relevant for inflation, $\phi \in [\phi_i ,\phi_e ]$. But the superpotential method by itself applies to other scalar field inflationary scenarios, as well, like, e.g., fast roll or ultra slow roll. The only condition for the method to be applicable without any modifications is that $\dot \phi$ should not change sign in the interval relevant for  inflation. 
In these generalized scenarios, however, it is more difficult to extract the asymptotic long wavelength behavior of the power spectra from the inflationary background field solutions. In the simplest case of leading order slow-roll, by using the time independence of the asymptotic power spectra, these may be related to the Hubble function and slow-roll parameters at horizon crossing by simple algebraic expressions like (\ref{Del-s}) and (\ref{Del-t}). In higher orders in slow-roll, algebraic expressions like
(\ref{Del-s}) and (\ref{Del-t}) may still be derived, but now involve higher orders (higher powers) of the slow-roll parameters at horizon crossing (see, e.g, \cite{ste-gong} for the second-order expressions). Beyond slow roll, the asymptotic power spectra must be extracted numerically from the numerical solutions of the Mukhanov-Sasaki equation, and simple algebraic expressions like (\ref{Del-s}) and (\ref{Del-t}) are no longer available. Obviously, the superpotential method shares this additional difficulty with all other approaches, but it still allows for a simple and efficient determination of the inflationary background solution which provides the starting point for the numerical solution of the Mukhanov-Sasaki equation. 

\section*{Acknowledgements}
The authors acknowledge financial support from the Ministry of Education, Culture, and Sports, Spain (Grant No. FPA2017-83814-P), the Xunta de Galicia (Grant No. INCITE09.296.035PR and Conselleria de Educacion), the Spanish Consolider-Ingenio 2010 Programme CPAN (CSD2007-00042), Maria de Maetzu Unit of Excellence MDM-2016-0692, and FEDER. CA thanks V. Vennin for very helpful discussions and J. Mas for help with Mathematica.

\appendix
\section{The Mukhanov-Sasaki equation in slow-roll inflation}
We want to solve the Mukhanov-Sasaki equation (\ref{MuSaEq}) under the assumption of slow-roll. Here by slow roll we mean that the Hubble flow parameters are (generically) nonzero, small and constant. That is to say, the slow roll parameters $\epsilon_n$ and their logarithmic time 
derivatives $\epsilon_{n+1} =H^{-1}\dot\epsilon_n/\epsilon_n$ are leading order (taken into account), whereas the time derivatives $\dot\epsilon_n \sim \epsilon_n\epsilon_{n+1}$ are next-to-leading order and may be ignored. 
Under this assumption, the term $z''/z$ in the Mukhanov-Sasaki equation has a very simple $\tau$ dependence. Up to first order in slow-roll, we calculate (using $z^2 = \kappa a^2 \epsilon$)
\be
\frac{z''}{z} \simeq (aH)^2 \left( 2 + \frac{3}{2} \epsilon_2 - \epsilon_1 \right) \simeq \frac{1}{\tau^2(1-\epsilon)^2} \left( 2 + \frac{3}{2} \epsilon_2 - \epsilon_1 \right) \simeq \frac{1}{\tau^2} \left( 2 + \frac{3}{2}\epsilon_2 + 3 \epsilon_1 \right) .
\ee
Further, we write up to first order
\be
2 + 3 \left( \frac{1}{2}\epsilon_2 + \epsilon_1\right) = 2 + 3(2\epsilon + \delta) \simeq \nu^2 - \frac{1}{4}
\ee
where $\nu$ is defined in (\ref{nu}). This is useful, because the resulting Mukhanov-Sasaki equation
\be
 v'' + \left(k^2 - \frac{\nu^2 -\frac{1}{4}}{\tau^2} \right) v =0
 \ee
has exact solutions in terms of the Hankel functions of the first and second kind. The correct limit for $|k\tau| >>1$ is provided by the Hankel function of the first kind, $H^{(1)}$, and the solution therefore reads
\be
v(k, \tau) = c \, \left( \frac{\pi |\tau|}{2}\right)^\frac{1}{2} H_\nu^{(1)} (|k\tau|) .
\ee
The asymptotic behavior of the Hankel function is
\be
\lim_{|k\tau| \to 0} H_{\nu}^{(1)}(|k\tau|) = \frac{i}{\pi} \Gamma(\nu) \left( \frac{|k\tau|}{2}\right)^{-\nu}
\ee
which, together with $\Gamma (\nu) \simeq \Gamma(\frac{3}{2}) = \sqrt{\pi}/2$, exactly reproduces the asymptotic expression (\ref{nu}) for $v^0(k,\tau)$.

\section{The linearised Eq. (\ref{eq-phi-only})}
In order to understand the behaviour of Eq. (\ref{eq-phi-only}) close to $\phi_0$ we linearise it, i.e., we insert $\phi (t) = \phi_0 + \delta \phi (t)$ and consider only the terms linear in $\delta \phi$. The resulting equation is
\be
\delta \ddot \phi + \sqrt{6\kappa V(\phi_0)} \delta \dot \phi + V_{,\phi\phi} (\phi_0) \delta \phi =0
\ee
where $V(\phi_0)$ and $V_{,\phi\phi} (\phi_0)$ are constants. This is the equation of a damped linear oscillator, which may be solved by the ansatz $\delta\phi = e^{\lambda t}$. The resulting algebraic equation for $\lambda$ is
\be
\lambda^2 + \sqrt{6\kappa V(\phi_0)} \lambda + V_{,\phi\phi} (\phi_0) =0
\ee
with the solutions
\be
\lambda = - \sqrt{\frac{3\kappa}{2} V(\phi_0)} \pm \sqrt{ \frac{3\kappa}{2} V(\phi_0) - V_{,\phi\phi} (\phi_0)} .
\ee
This leads to two real, negative roots $\lambda_\pm = - |\lambda_\pm |$ precisely for potentials which obey the "Breitenlehner-Freedman bound" (\ref{SkT-bound-V}). The resulting solution $\delta \phi$ is exponentially decaying ("overdamped oscillator"),
\be
\delta \phi = a e^{-|\lambda_+| t} + be^{-|\lambda_-|t}
\ee
and cannot reach $\delta \phi =0$ in a finite time, as announced. On the other hand, for potentials violating the bound (\ref{SkT-bound-V}), the two roots $\lambda_\pm$ are complex and lead to an oscillating, damped $\delta \phi$ (underdamped oscillator). Such oscillations produce a specific imprint on certain inflationary observables (oscillatory behaviour in certain non-Gaussian spectra). It was concluded in \cite{ACDP} that the observation of these oscillatory patterns would exclude all inflationary models based on the superpotential (or Hamilton-Jacobi) equation, even in the multifield case. As pointed out already in the main text, we want to emphasize again that this argument only excludes the {\em global} existence of a superpotential, but not necessarily the existence of local solutions of the superpotential equation in some regions of field space, which, in general, is sufficient for our purposes.


\end{document}